\documentclass[acmsmall, manuscript]{acmart}
\AtBeginDocument{%
  }

\usepackage{xcolor}
\usepackage{ragged2e}
\usepackage{longtable}
\usepackage[utf8]{inputenc}
\usepackage{array}
\usepackage[noautocite]{annotation}

\acmConference[CHI PLAY '26]{The Annual Symposium on Computer-Human Interaction in Play}{November 02--05,
  2026}{York, GB}

\begin{document}

\title[The Landscape of Misinformation Literacy Games: A Systematic Mapping Review of Game Designs and ...]{The Landscape of Misinformation Literacy Games: A Systematic Mapping Review of Game Designs and Evaluation Practices}

\author{Omed Abed}
\authornote{Both authors contributed equally to this work.}
\affiliation{%
  \institution{Rhine-Waal University of Applied Sciences and Doctoral School NRW, Department of Computer and Data Science}
 \city{Kamp-Lintfort}
  \country{Germany}
}
\email{omed.abed@hochschule-rhein-waal.de}
\orcid{0009-0004-8989-9514}

\author{Smi Hinterreiter}
\authornotemark[1]
\affiliation{%
  \institution{University of Würzburg}
\country{Germany}
 \city{Würzburg}
}
\email{smi.hinterreiter@uni-wuerzburg.de}
\orcid{0000-0002-7029-2753}

\author{Sijia Guo}
\affiliation{%
  \institution{RWTH Aachen University}
 \city{Achen}
\country{Germany}
}
\email{sijia.guo@rwth-aachen.de}
\orcid{0009-0008-7656-356X}

\author{Isao Echizen}
\affiliation{%
  \institution{National Institute of Informatics}
 \city{Tokyo}
\country{Japan}
  }
\email{iechizen@nii.ac.jp}
\orcid{0000-0003-4908-1860}

\author{Timo Spinde}
\affiliation{%
  \institution{University of Göttingen}
 \city{Göttingen}
\country{Germany}
  }
\email{t.spinde@media-bias-research.org}
\orcid{0000-0003-3471-4127}

\author{Matteo Große-Kampmann}
\affiliation{%
  \institution{Rhine-Waal University of Applied Sciences and Doctoral School NRW, Department of Computer and Data Science}
   \city{Kamp-Lintfort}
  \country{Germany}
}
\email{matteo.grosse-kampmann@hochschule-rhein-waal.de}
\orcid{0000-0001-9127-969X}

\renewcommand{\shortauthors}{Abed \& Hinterreiter et al.}

\begin{abstract}
    This systematic mapping review examines 54 gamified interventions aimed at combating misinformation. We analyze the theoretical frameworks, game design, and media contexts of academic and non-academic games. The results indicate a field dominated by web-based simulations and trivia formats. While psychological inoculation is a prominent theoretical framework in the field, most interventions prioritize content-discernment tasks over proactive prebunking strategies. 
    We identify systematic gaps in the identified games. Current interventions remain largely text-centric, focusing on news articles and text-based content, despite the growing use of audiovisual content (e.g., short videos) in the digital landscape. We conclude our study by discussing implications for addressing the evolving, multimodal nature of potentially misleading content using gamified interventions. 
\end{abstract}


\setcopyright{cc}
\setcctype{by}
\acmJournal{PACMHCI}
\acmYear{2026} \acmVolume{10} \acmNumber{7} \acmArticle{GAMES068}
\acmMonth{11} \acmDOI{10.1145/3831350}




\maketitle
 \AddAnnotationRef

\section{Introduction}
    A current societal challenge is the increase in misleading information and narratives \cite{Velichety2022_FakeNews}, particularly through short posts on social media that contain images or videos \cite{turos-2024}.
    Content containing misinformation or disinformation can influence opinions, manipulate audiences, prevent the formation of truthful beliefs \cite{harris-2023}, discourage democratic engagement, and weaken trust in news \cite{Ognyanova2020, Sardo2024} as well as societal cohesion by deepening polarization \cite{Lilja2024, Surjatmodjo2024}.
    Misinformation is defined as "false information that is shared without the intention to mislead or cause harm" \cite[Table 3]{Ameur2023}.
    Exposure to harmful health misinformation has been associated with confusion, reduced trust, and risky behaviors among younger users, e.g., injecting bleach to cure diseases \cite{Beran2019, GrosseKampmann2025}.
    Disinformation intentionally misleads and is often politically motivated, e.g., through the manipulation of elections \cite{vaccari-2024, mueller2019reportvol1, benkler-2018}, or aims to change public opinion on topics such as support for Ukraine \cite{DFRLab2024_PolishFarmers}.
    Besides social media platforms, legacy media can also play a significant role in amplifying or disseminating false narratives \cite{Tsfati02042020}.
    Often, the umbrella term "fake news" is used to refer to false information \cite{Ameur2023}.
    Young adults are especially at risk of manipulation, as they primarily consume information online through user-generated content platforms, which are channels where false information and manipulative narratives can spread quickly~\cite{EuropeanParliament2025}.
    One study reports that almost 20 percent of the top 20 search results on TikTok contain misinformation about major news topics~\cite{Pira2023DisinformationAP}. Users struggle to identify misleading content, which they encounter at least weekly in their feeds~\cite{Hoffman2023TheIO}.
    
    To mitigate these adverse effects, media literacy, specifically the capacity to evaluate the credibility of online information and distinguish accurate content from misinformation, has become an essential competency for digital media users \cite{Buckingham2006_MediaLiteracy, NetwerkMediawijsheid2020_CompetencyModel}.
    However, the delivery of media literacy education to children, youth, and young adults depends on individual instructors, as formal school curricula rarely integrate it systematically \cite{Schmitz2024_MediaLiteracyEducation}.
    Here, serious games offer a scalable and engaging approach to teach critical thinking and information verification strategies, leveraging the widespread acceptance of digital games \cite{susiSeriousGamesOverview2007, basol-2020}.

    Prior work suggests that game-based interventions can improve misinformation resilience, often by having players practice or "prebunk" manipulation strategies (e.g., \emph{Bad News}) \cite{Maertens2021}.
    At the same time, many well-studied interventions are guided, choice-based simulations that emphasize short, text-forward interactions \cite{Maertens2021, Rev1_2024_Tackling}.
    This focus is increasingly challenged by the prominence of multimodal misinformation, including persuasive short-form videos and image-based formats \cite{Jiang2024}.
    Additionally, some media and information literacy interventions can reduce trust in misinformation while also inadvertently lowering trust in accurate news \cite{MadridMorales2024}.

    Prior reviews have already advanced our understanding of misinformation games, but they tend to focus on relatively small corpora, short-term evaluations, and a narrow set of mostly text-based social media simulations \cite{Rev1_2024_Tackling, Rev2_2023_EvaluatingAsTools, Rev3_2024_LiteracyAtPlay, Rev4_2023_PlayingFakeState}.
    They also differ in scope and analytic emphasis, with some focusing primarily on learning outcomes \cite{Rev2_2023_EvaluatingAsTools} and others on game design or intervention types \cite{Rev1_2024_Tackling}.
    As a result, we still lack a broader view of how misinformation education games are designed, what pedagogical strategies they use, how they connect learning objectives to game mechanics, which media environments they address, and how they are studied in practice.
    In addition, publicly released games developed outside academia (e.g., by NGOs, journalists, studios, or civic education initiatives) are often excluded because they lack peer-reviewed evaluations, even though they may represent more fully realized game experiences (e.g., \emph{Wiebkes Wirre Welt}, \emph{Influence Inc.}).

    To address these gaps, we conduct a systematic mapping review that combines a PRISMA-guided literature search with a qualitative search for publicly available misinformation education games.
    We differentiate between academic games, defined as games accompanied by a corresponding peer-reviewed publication, and non-academic games, which lack such publications and may include both commercial titles and freely available educational games.
    Together, these sources allow us to map the current landscape of misinformation education games across game design, pedagogical approaches, media context, audience, and evaluation.
    Against this background, our review is guided by the research questions:
    \begin{itemize}
        \item \textit{(RQ1)} What types of gamified interventions have been developed to address misinformation?
        \item \textit{(RQ2)} What teaching objectives and theories do they employ?
        \item \textit{(RQ3)} What game designs and formats are used?
        \item \textit{(RQ4)} What media context and content does the game cover?
        \item \textit{(RQ5)} What populations and study contexts are used?
        \item \textit{(RQ6)} What are the measured metrics, results, and key findings?
    \end{itemize}

    Our review extends prior work in both scope and coverage, including 54 games and 31 papers.
    Overall, 30 games have peer-reviewed corresponding papers; 17 apply quantitative learning outcome measures and 9 apply qualitative or mixed-method measures; 4 papers present primarily the game concept or analyze the content and mechanics of the game.
    In addition, we broaden the perspective of academic game research by including 24 non-academic games and systematically quantifying their designs.

    By mapping and thematically analyzing both academically evaluated interventions and publicly released games across categories such as learning objectives, theoretical grounding, game mechanics, and player roles, we map the games of this review into five types: Verification Games, Strategy Narrative, Platform Simulations, Role Immersion, and Systems-Power Games.
    Our goal is to inform the design of future misinformation education games that are not only effective but also more gameful and engaging, and that better reflect the multimodal media environments where misinformation is encountered today.

\section{Background}
\label{sec:related_work}
    The contemporary digital environment is characterized by low barriers to creating and publishing content \cite{wardle2017}. Moreover, this issue is  driven by algorithmic amplification, which prioritizes sensationalism over accuracy \cite{vosoughi-2018}. Within this ecosystem, information disorder can spread quickly.
    Wardle et al. categorized information disorder into three groups: (1) Misinformation, false information shared without the intent to mislead, (2) Disinformation, information created with the intent to deceive, and (3) Malinformation, the use of truthful information in a malicious context to cause harm \cite{wardle2017}. This manipulates audiences, prevents the formation of truthful beliefs, and weakens trust in information sources generally \cite{harris-2023, Ognyanova2020}.

    The use of Generative AI, such as image generators and chatbots like ChatGPT, Gemini, or Grok, has amplified this threat by enabling the rapid production of multimodal, highly realistic synthetic media \cite{lopez2025}. Bad actors can use 
    synthetic media containing misleading, false, or harmful content to drive counternarratives that even dismiss truthful evidence as fake \cite{SCHIFF_BUENO_2025}.
    Consequently, this increases the cognitive burden on users who must verify not only the content but also its origin \cite{SCHIFF_BUENO_2025, lopez2025}, necessitating high levels of media literacy.

  \subsection{Media Literacy and Misinformation Education}
    One established approach to media literacy is civic online reasoning (COR). Media literacy curricula \cite{aufderheideMediaLiteracyReport1993a} and COR interventions \cite{mcgrewLearningEvaluateIntervention2020, axelssonLearningHowSeparate2021} aim to strengthen individuals’ veracity discernment abilities, the ability to critically evaluate whether information is credible or misleading. These approaches typically provide structured guidance to support systematic source evaluation and verification practices.
    For example, media literacy frameworks often draw on structured evaluation strategies such as the CRAAP test (Currency, Relevance, Accuracy, Authority, Purpose) \cite{craapBlakeslee2004} or the SIFT method (Stop; Investigate the source; Find better coverage; Trace claims, quotes, and media to the original context) \cite{caulfield2023verified}. These heuristics provide step-by-step procedures for assessing the credibility of online information. Similarly, COR interventions often rely on interactive checklists \cite{GlassmanWHO2022} and web-based learning environments to operationalize such strategies. For instance, the interactive tool \emph{The News Evaluator} provides users with structured tutorials, hands-on tasks, and both implicit and explicit feedback to support critical engagement with online content~\cite{axelssonLearningHowSeparate2021}. By packaging COR objectives into guided activities, these interventions promote systematic evaluation strategies.

    A related framework is inoculation theory, originally proposed by \citet{mcguire1960}. Drawing on a medical analogy, the theory posits that exposing individuals to a weakened form of misinformation can build resistance to future persuasive attempts \cite{Maertens2021}. In the context of misinformation, this approach typically involves forewarning individuals about an impending persuasive threat and providing refutational preemption. This process equips users with cognitive strategies to recognize and counter common manipulation techniques, such as emotional appeals, polarization, or impersonation.

 \subsection{Misinformation Education through Games}
    Serious games and gamification can serve as effective pedagogical tools for teaching information evaluation skills and enhancing educational settings \cite{oberdorferBetterLearningGaming2021, GroveSeriousGames2010, VanEckGBL2006, GeeGamesLearningLiteracy2003, GreitzerCognitiveScienceForGamification2007}. Gamification refers to the integration of game design elements into non-game contexts \cite{DeterdingGamificationSeriousGames2011, krathRevealingTheoreticalBasis2021}, whereas serious games are developed with a primary purpose beyond entertainment, often education \cite{susiSeriousGamesOverview2007}.
    Gamification and serious games sustain attention through active participation, helping complex content reach audiences who might otherwise disengage. Replacing passive consumption with interactive experiences also supports longer-term retention of learning objectives \cite{Maertens2021}.

    Beyond increasing engagement through interaction \cite{Rev2_2023_EvaluatingAsTools}, serious games embed learning objectives within their mechanics and narratives \cite{arnabMappingLearningGame2015}.
    Recent reviews of design frameworks identified the systematic alignment of game design with educational objectives as a central determinant of effectiveness \cite{maxim-2024}. The Gamified Knowledge Encoding (GKE) model describes this link as learning content that is segmented into rules that are encoded into interactive game mechanics. The repeated executions of those mechanics during play then build a mental model that can transfer to real-world situations \cite{oberdorferGamifiedKnowledgeEncoding2018}. Beyond mechanics, the narrative framing and aesthetic presentation of a game are themselves part of how it communicates its content rather than a neutral wrapper around its mechanics, influencing which competencies it can convey \cite{Rev3_2024_LiteracyAtPlay}.
        
\begin{figure*}[h]
            \centering
            \includegraphics[width=0.5\linewidth]{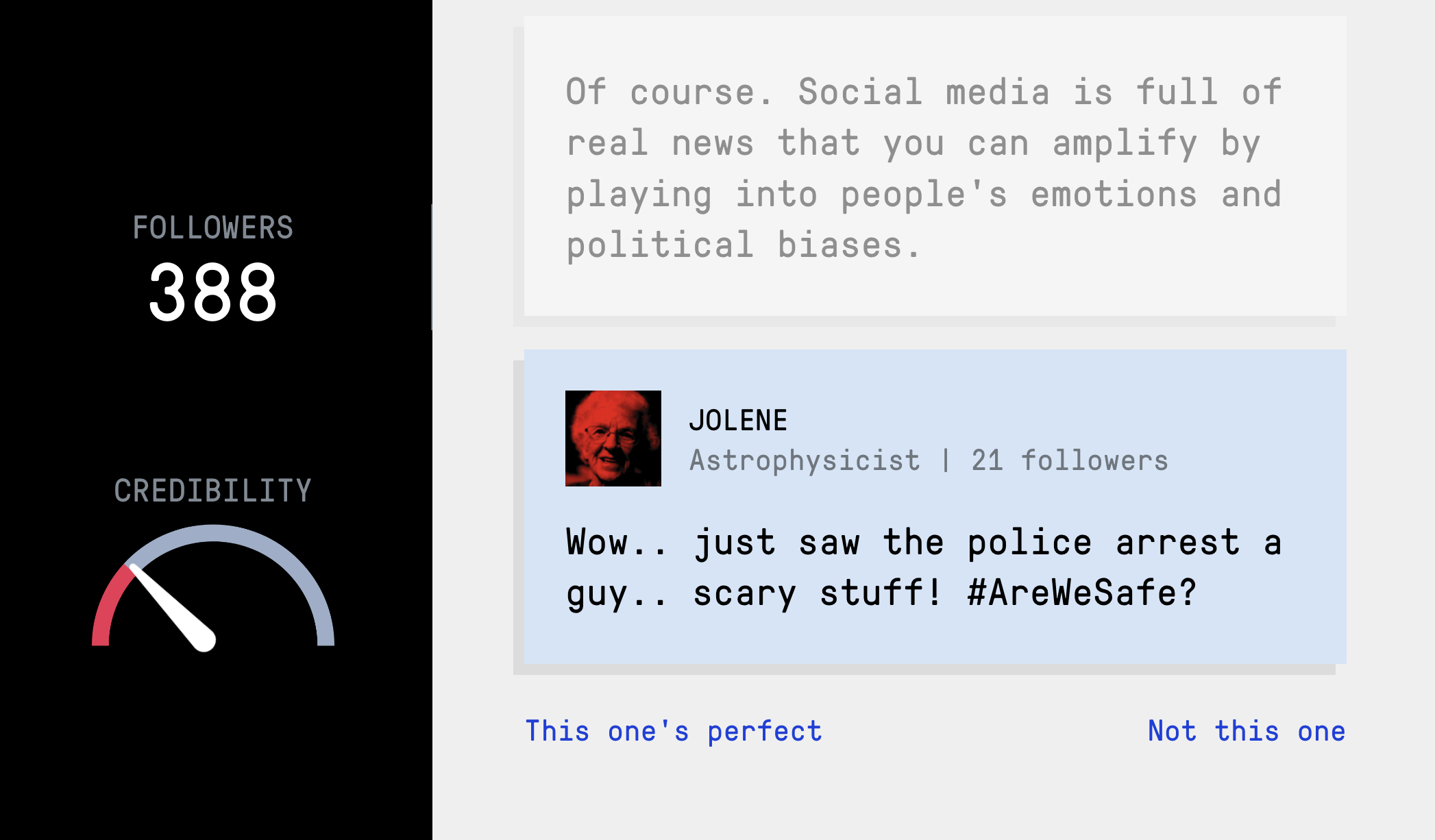}
            \caption{\emph{Bad News} (released in 2018, \cite{roozenbeek-2019, basol-2020}), a social-platform simulation game, in which players act as a bad actor. The game inoculates players against misinformation by having players use six manipulation strategies: Impersonation, Emotion, Polarization, Conspiracy, Discredit, and Trolling.}
            \Description{Screenshot of the game Bad News. A left panel shows a follower count of 388 and a credibility gauge. A right panel shows supporting text about amplifying emotionally, politically biased content, and a simulated social media post by a user named Jolene with the options "This one's perfect" and "Not this one".}
            \label{fig:simulationgamesbadnews}
        \end{figure*}

    Gamified interventions have become a central application of inoculation theory through preemptive and active inoculation. Unlike passive educational approaches, games such as \emph{Bad News} (Figure~\ref{fig:simulationgamesbadnews}, \cite{roozenbeek-2019}) allow players to assume the role of a disinformation producer. By creating misleading content within a simulated, text-based social media environment\footnote{a text-based social media environment such as the social media platform \emph{X}}, players engage directly with common manipulation strategies. Conversely, in \emph{Misinformation is Contagious} \cite{barzilai-2025}, players take the role of the good producer and try to avoid publishing misinformation. This active involvement supports the internalization of deceptive techniques and improves the ability to recognize similar strategies in real-world contexts \cite{Roozenbeek2019, Rev4_2023_PlayingFakeState}. Longitudinal evidence suggests that \emph{Bad News} can increase resilience to misinformation for up to three months \cite{Maertens2021}.
    However, most existing interventions rely on text-based simulations and pay limited attention to contemporary formats such as short-form video content \cite{Jiang2024}. This points to a gap between a media landscape shaped by videos and the predominantly text-based design of educational misinformation games.

    \section{Related Work}
    \label{sec:existing_lit_revs}   
    Gamified misinformation interventions have been mapped by several reviews (see Table~\ref{tab:comp_prior_rev}) \cite{Rev1_2024_Tackling, Rev2_2023_EvaluatingAsTools, Rev3_2024_LiteracyAtPlay, Rev4_2023_PlayingFakeState}, identifying a reliance on short-term, small-scale evaluations. \citet{Rev1_2024_Tackling} conducted a systematic review of 15 papers (covering 12 unique games) and found that studies typically featured a median of $n=196$ participants with a median playing time of only 15 minutes. Most games used preemptive inoculation strategies. Furthermore, only two of the analyzed studies involved minors, suggesting a significant gap in researching younger, more vulnerable demographics. Similarly, the rigor of evaluation varied, with only eight studies using a formal control group and seven employing a pre/post-test design \citep{Rev1_2024_Tackling}.

    Across recent reviews, digital interventions consistently exhibit limited diversity in both format and gameplay structure. \citet{Rev4_2023_PlayingFakeState} analyzed 22 games, identifying a dominance of browser-based \textit{Quiz games} ($n=12$) and \textit{Text Adventures} ($n=6$), with playtimes often averaging only five minutes. This genre homogeneity is echoed by \citet{Rev3_2024_LiteracyAtPlay}, who conducted a thematic analysis of 100 media literacy games. 
    Their findings indicate that most interventions are single-player and strictly linear, primarily focusing on misinformation and privacy rather than other media competencies.

    The mapping of mechanics to educational goals remains a challenge. \citet{Rev2_2023_EvaluatingAsTools} analyzed 24 games, including commercial games sourced from Steam, and developed a point-based system to evaluate learning outcomes such as understanding algorithmic effects and source credibility. Despite including commercial games, most interventions that cover learning objectives still prioritize simple mechanics, such as asking players to discern news veracity or spreading fake news to understand its mechanics and consequences. However, \citet{Rev2_2023_EvaluatingAsTools}'s focus on Steam excluded a range of web-based, non-academic misinformation games.

\setlength{\tabcolsep}{3pt}
\begin{table}[h]
\centering
\caption{Comparison of coding dimensions across prior reviews and the present work. Dashes indicate that the variable was not in the review as it was not in the scope of their research questions. N/A indicates that the variable could have been coded but was not included.}
\label{tab:comp_prior_rev}
\small
\begin{tabular}{@{}p{2.3cm} p{2cm} p{2cm} p{2cm} p{2cm} p{2cm}@{}}
\toprule
\textbf{Variable} & \textbf{Kiili et al.\ \cite{Rev1_2024_Tackling}} & \textbf{DeJong \cite{Rev4_2023_PlayingFakeState}} & \textbf{Glas et al.\ \cite{Rev3_2024_LiteracyAtPlay}} & \textbf{Contreras \&Eguia\cite{Rev2_2023_EvaluatingAsTools}} & \textbf{Our review} \\  
\midrule  
Scope & Empirical studies; critical reading focused & Digital fake news games &  Digital media-literacy games & Digital media-literacy games incl. commercial titles & Digital + analog games; academic and non-academic \\\hline
Games (Paper) & 12 (15) & 22 games & 53 played, 12 in-depth & 24 & 54 (30) \\\hline
Non-academic games & -- & -- & -- & \checkmark & \checkmark \\\hline
Game type / genre & \checkmark & \checkmark  & \checkmark & \checkmark & \checkmark \\\hline
Game mechanics & \checkmark (descriptive) & -- & partial & \checkmark & \checkmark \\\hline
Player role$^{2}$ & \checkmark (2 categories) & \checkmark (2 categories) & partial (3 categories) & -- & \checkmark (7 categories) \\\hline
Media context & \checkmark & -- & -- & -- & \checkmark \\\hline
Modality$^{1}$ & -- & -- & -- & -- & \checkmark \\\hline
Theo. framework$^{2}$ & \checkmark & -- & Competency Mapping & Curriculum mapping & \checkmark \\\hline
Learning\newline objectives  & partial & partial & \checkmark & \checkmark & \checkmark \\\hline
Target audience / age & \checkmark & \checkmark & -- & \checkmark & \checkmark \\\hline
Playing time & \checkmark & \checkmark & -- & \checkmark & -- \\\hline
Availability\newline/platform & -- & \checkmark & \checkmark & \checkmark & partial \\\hline
Thematic analysis$^{2}$& -- & -- & \checkmark (deductive, MIL) & -- & \checkmark (in- + deductive) \\\hline
\toprule
\multicolumn{6}{@{}l}{\textit{Paper-level dimensions (coded only for games with empirical studies)}} \\\hline
UX/PX measures$^{1}$ & -- & N/A & N/A & N/A & \checkmark (30-game subset) \\\hline
Learning\newline outcomes & \checkmark & N/A & N/A & N/A & \checkmark (30-game subset) \\\hline
Sci. impact (FWCI) & \checkmark & N/A & N/A & N/A & -- \\\hline
\multicolumn{6}{p{0.9\linewidth}}{\footnotesize $^{1}$ Dimension introduced in the present work; all other dimensions were coded in at least one prior review. $^{2}$ Where a dimension is shared, cell entries show differences in granularity (e.g., player role in seven categories here versus two or three in prior reviews).}\\
\bottomrule
\end{tabular}
\end{table}

    The identified reviews except \citet{Rev2_2023_EvaluatingAsTools} examine only a limited subset of misinformation games and exclude games developed for educational purposes that have no corresponding publication. Examples such as \emph{Wiebkes Wirre Welt} ("Wiebkes tangled world") and \emph{Influence Inc.}\footnote{\emph{Influence Inc.} is the commercial successor of \emph{Fake it to Make it} \cite{urban-2018}.} incorporate more advanced gameplay, content, and design elements but are left out of academic reviews.

    Table~\ref{tab:comp_prior_rev} maps the variables used in each of the prior reviews and in the present work. Three patterns show. First, no prior review combines game-level and paper-level coding at scale. While Kiili et al. \cite{Rev1_2024_Tackling} analyzes empirical studies, their corpus is limited to 12-games. DeJong \cite{Rev4_2023_PlayingFakeState}, Glas et al. \cite{Rev3_2024_LiteracyAtPlay}, and Contreras-Espinosa \& Eguia-Gomez \cite{Rev2_2023_EvaluatingAsTools} engage with game designs across larger samples without coding corresponding studies.
    Second, several dimensions are not addressed by any prior review, such as the modality of content presented in the game and whether the empirical studies measure player experience. 
    Third, prior reviews mostly exclude publicly released non-academic interventions.

    This study addresses these limitations by examining a substantially larger corpus of misinformation games ($n=54$), more than double the scope of comparable recent reviews \cite{Rev1_2024_Tackling, Rev2_2023_EvaluatingAsTools, Rev4_2023_PlayingFakeState}. It also broadens the sample's diversity by including both academically evaluated games and publicly released interventions. We systematically compare educational strategies, game types, narratives, media contexts, covered modalities, target audiences, and reported outcomes. By mapping this broader landscape, the review identifies trends, game types, and design patterns and explores how academically developed games and publicly released, non-commercial games can inform and strengthen each other’s approaches.

\section{Methodology}
\begin{figure*}[h]
            \centering
            \includegraphics[width=0.8\linewidth]{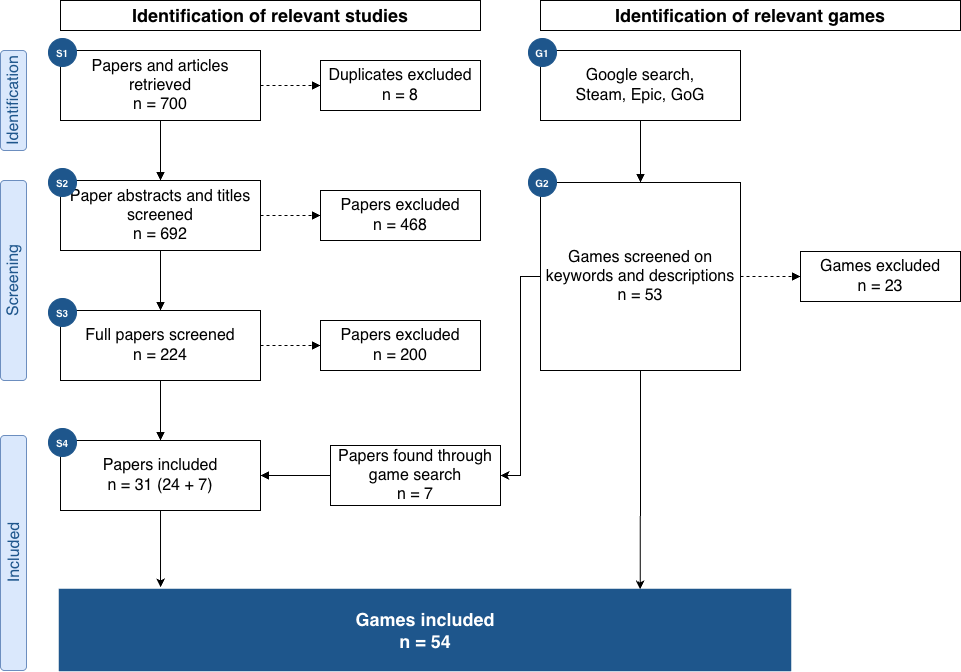}
            \caption{Game Selection and Screening Process. Steps are labeled S1-S4 (structured literature search) and G1-G2 (Game search)}
            \Description{Two-track flow diagram showing the selection and screening process for studies (left) and games (right), organized by the stages Identification, Screening, and Included. Left track (studies): 700 papers and articles retrieved; 8 duplicates excluded; 692 titles and abstracts screened, with 468 papers excluded; 224 full papers screened, with 200 papers excluded; 24 papers included. Right track (games): games identified via Google search and game platforms (Steam, Epic, and GOG); 53 games screened based on keywords and descriptions, with 23 games excluded; the final output is 54 games included. A cross-link indicates that 7 papers were additionally found through the game search and fed into the paper screening/inclusion process.}
            \label{fig:game_selection}
        \end{figure*}

    The objective of this mapping review \cite{GrantMapping2009} is to systematically identify, analyze, and map gamified misinformation interventions designed. The review follows the PRISMA 2020 guidelines for transparent reporting of search, screening, and inclusion procedures \cite{Page2021_PRISMA} detailed in Figure~\ref{fig:game_selection}. In addition to a formal academic database search, we conducted a qualitative search on Google and game platforms to identify games not represented in scholarly databases.

    \label{search strategy}

    \subsection{Search Strategy}
    \subsubsection{Literature Search}
        \label{literature_search}
        A systematic literature search was conducted to identify relevant scientific literature. Based on the research questions, two keyword groups were developed (see Table~\ref{tab:scholar-queries}). The first group contains keywords about the problem domain, while the second group defines the intervention types. Boolean operators (AND/OR) were used to systematically combine these keyword groups into seven distinct search strings. The "OR" operator was used to include synonyms, while the "AND" operator was used to intersect the problem domain with the intervention types. The search was executed on Google Scholar on September 1, 2025, using a script to retrieve the articles, and returned 700 entries. Google Scholar served as the primary database as it has great coverage, broad indexing, and a high sensitivity to return relevant research \cite{NashGoogle26}.

    \subsubsection{Game Search}
        \label{game search}
        Additionally, an exploratory search for educational and serious games was conducted on Google. This search aimed to identify games that are not evaluated in academic papers but are relevant to the landscape of misinformation interventions. We conducted Google searches using the terms "fake news game", "misinformation game", "disinformation game", and "desinformation game". We queried the game distribution platform Steam with the keywords "fake news", "misinformation", "disinformation", "desinformation", and "news". Using the snowball method, we identified additional relevant games. Lastly, game titles found through Google were searched for in Google Scholar.\footnote{If the game title did not yield hits, the search term was extended to include "game" or "misinformation".} If a corresponding paper was found, we added it to the corpus of scientific games.

    \subsection{Selection Process}
        As this work combines a systematic literature search and qualitative searches, we describe both selection processes below. Throughout both processes, two independent coders applied the inclusion criteria. Disagreements were resolved through discussion with a third coder. The final corpus includes all games from both searches.

        \textbf{Systematic Literature Search and Inclusion Criteria.}
        The PRISMA search returned 700 records. After removal of duplicates (n = 8), the corpus contained 692 unique records. Article relevance was assessed in a two‑stage screening process. First, records were excluded at title and abstract level if they did not meet all of the following inclusion criteria: (1) the focus is on misinformation, disinformation, media literacy, or veracity discernment; (2) the work presents a game or gamified educational intervention; (3) the publication language is English or German; and (4) the work is not a literature review. Second, the remaining articles underwent full‑text screening and were included if (5) the paper described the game design or the game was available to play, and (6) the gameplay contained educational elements related to information-veracity discernment. We included $n=31$ entries for $n=30$ games listed in Table~\ref{tab:rev_papers}, including $n=7$ games identified via the supplementary search with corresponding publications.

        \textbf{Qualitative Game Search and Inclusion Criteria.}
        Games identified via our supplementary search were included if they (1) were available to play, (2) contained relevant educational content for veracity discernment, (3) had gameplay focused on veracity discernment, and (4) had gamification mechanics exceeding a multiple-choice quiz. Each game is analyzed by two coders ($\kappa_{\text{linear}} = .75$; $\kappa_{\text{quadratic}} = .82$). From $ n= 58$ found games, we included $ n=24$, excluding the aforementioned $n=7$ games with corresponding papers (Google/Included $ n= 24/15$, Snowball/Included $ n=22/8 $, Steam/Included $ n= 7/2$, Paper Snowball/Included $ n= 5/5$).

        \textbf{Final Corpus.} The final corpus contains 54 games. A comparison against the four prior reviews  \cite{Rev1_2024_Tackling, Rev2_2023_EvaluatingAsTools, Rev3_2024_LiteracyAtPlay, Rev4_2023_PlayingFakeState} shows that 30 (56\%) of the 54 games are unique entries that have not been analyzed in any prior review, while 24 (44\%) appear in at least one (see Table~\ref{tab:RQ1,RQ3,RQ4}).

    \subsection{Analysis and Coding of Games}
    Each coder first familiarized themselves with the games by playing them and reviewing available supporting material. Observable variables such as platform, game type, and media context were coded in parallel. More interpretive variables, including game mechanics, teaching objectives, story, and player role, were documented through analytic notes and then refined through iterative comparison across cases. Ambiguous cases were discussed with an additional coder. All final labels are reported in Table~\ref{tab:labels}. Game type, mechanics, player role, and narrative framing were developed inductively during the iterative comparison of the games. Learning-objectives and learning-theories were derived deductively from established media-literacy frameworks and instructional theories (e.g., COR, SIFT, CRAAP; see~\ref{sec:related_work}).
    Beyond categorical labeling, coders also noted qualitative observations such as the salience or centrality of certain elements (e.g., relative mechanic complexity) which informed the thematic analysis without being captured in the structured coding labels.
    In addition, one researcher reviewed the corresponding paper to summarize the study design, sample, and reported outcomes (RQ5, RQ6).

    To identify common themes and types of gamified interventions for misinformation education (RQ1), a second-cycle thematic coding process was used to group recurring patterns across mechanics, narrative, player role, teaching objectives, and theoretical framing into broader themes according to shared pedagogical objectives and game designs. We focused on connecting game mechanics with teaching objectives, following the GKE model \cite{oberdorferGamifiedKnowledgeEncoding2018}.
    This inductive-deductive process produced higher-order thematic categories that describe how misinformation games position players, structure learning, and use game design and mechanics to teach about misinformation. The purpose of this step was not to produce a rigid typology, but to identify recurring patterns that summarize the corpus.

    \begin{table}[h]
\centering
\caption{Coding Labels}
\label{tab:labels}
\small
\begin{tabular}{@{}p{0.16\linewidth} p{0.76\linewidth} >{\raggedleft\arraybackslash}p{0.04\linewidth}@{}}
\toprule
\textbf{Variable} & \textbf{Labels} & \textbf{RQ} \\
\midrule
        Topic & (Mis)Information Literacy; Conspiracy Theories; Propaganda; Disinformation; Manipulation Strategies; Trolls, Satire, Humor, Ads; Bots; Social Media; Polarization & RQ1\\\hline
        Intervention Type & Video Game; Card Game; Board Game; Classroom Game; AR; Escape Room; TTRPG & RQ1\\\hline
        Framework / Learning Objective & Detect misleading/truthful content; Social Media Literacy; Procedural literacy; Socio- emotional awareness; Normative behavior; System/power understanding; Strategy awareness & RQ2\\\hline
        Learning Theory & Procedural; Declarative knowledge transfer; Inoculation/prebunking; Procedural strategy use; Constructivism/ situated learning; Corroboration; Veracity discernment; Source evaluation; Bias recognition; Metacognitive; Heuristic vs.\ analytical; Narrative-based & RQ2 \\\hline
        Game Type & Choice-based Simulation; Trivia; Classroom; Board Game; Card Game; Decision Tree; Other; Point and Click; RPG; Simulation & RQ3\\\hline
        Game Mechanics & Source/ content evaluation; Headline classification; Scoring/feedback loops; Social-media simulation; Narrative framing (none, low to medium, strong); Content creation; Sharing decision; Story decision; Quiz/ Trivia; Competitive; Cooperative; role-play; Progression & RQ3\\\hline 
        Player Role & Bad Actor; Good Actor; Detached Evaluator; Strategic Actor; Embedded Evaluator; Accidental Bad Actor; Opponent& RQ3\\\hline
        Platform & Web-Based; Mobile; Card Game; Board Game; Desktop; Console; Other & RQ3\\\hline
        Media Context & News Articles; Social Media; Multi-modal; Video; Other; No specific media context & RQ4\\
        \bottomrule
    \end{tabular}
\end{table}

    The corpus was analyzed using a mixed quantitative and qualitative content analysis. The analysis focused on game properties relevant to the research questions and could be compared across cases, including platform, game type, mechanics, narrative, topic, player role, teaching objectives, intervention type, theoretical framework, and the kind of misinformation represented. These variables were chosen because they capture both the design of the game and its educational framing: platform and game type describe the medium and format (RQ3); mechanics, narrative, and player role describe the game design and how misinformation learning is implemented in it; topic and misinformation type describe the learning material (RQ4); and teaching objectives, intervention type, and theoretical framework capture the pedagogical grounding of the game (RQ2).
    Most variables in Table~\ref{tab:labels} were coded categorically for presence or type. Narrative was the exception. Given its centrality to several games' pedagogical design, it was additionally coded for strength/integration level. We rated each game's \emph{narrative strength} on an ordinal scale (none, weak, medium, strong) to reflect how a fictional storyline shapes gameplay and learning. 
    \emph{None} denotes games with no storyline, such as classification or quiz games without a setting or character; the player acts as a detached evaluator;
    \emph{weak} a simple, linear storyline which sets the narrative of the game but does not develop over the course of play, with limited replayability;
    \emph{medium} a more developed setting or role that structures gameplay;
    \emph{strong} a game in which players are embedded within a fictional world and their decisions carry narrative consequences, such that learning is transported through the narrative.

\section{Results}
In the following, we first report numbers on the types of misinformation games, their design characteristics, mechanics, and media contexts, as well as their modalities, theoretical frameworks, and learning objectives of the 54 games. Afterward, we describe the five themes identified through the thematic analysis. Finally, we examine how these games have been empirically evaluated and which samples and outcome measures have been reported.

\subsection{Mapping Review of Misinformation Games} \label{sec:mapping}
\subsubsection{Topic and Platform of Games (RQ1).}
Our quantitative analysis of 54 games shows a dominance of digital games. Video games make up the majority of the corpus ($n = 41$, 76\%), consisting mostly of web-based games ($n=32$, 59\%). In contrast, other digital formats are scarce. Mobile applications ($n=3$), mobile and web-based ($n=2$), and dedicated desktop or console releases ($n=6$) represent only a small part. The remaining titles comprise analog and hybrid formats, including card games ($n=5$), board games ($n=2$), and niche interventions such as classroom games ($n=4$), escape rooms ($n=2$), tabletop games ($n=2$), and augmented reality games ($n=1$).
The proportions indicate that misinformation games are predominantly built for individual or small‑group screen‑based play, with a noticeable niche of table‑top and classroom implementations.

Most games ($n=49$, 91\%) addressed misinformation literacy in general. Twelve games were explicitly focused on social media, though many others used it implicitly within the game context. Nine games emphasized manipulation strategies. Five games each addressed conspiracy theories, propaganda, trolls/satire/humor. Four games focused on disinformation, and four addressed advertising. Three games included bots as a topic, such as \emph{Harmony Square}, \emph{Escape Fake}, which feature bots that create traffic and false consensus, and \emph{Spot The Troll} turning bot detection into the core task.

\subsubsection {Frameworks and Learning Objectives (RQ2)}

\begin{figure*}[h]
            \centering
            \includegraphics[width=0.7\linewidth]{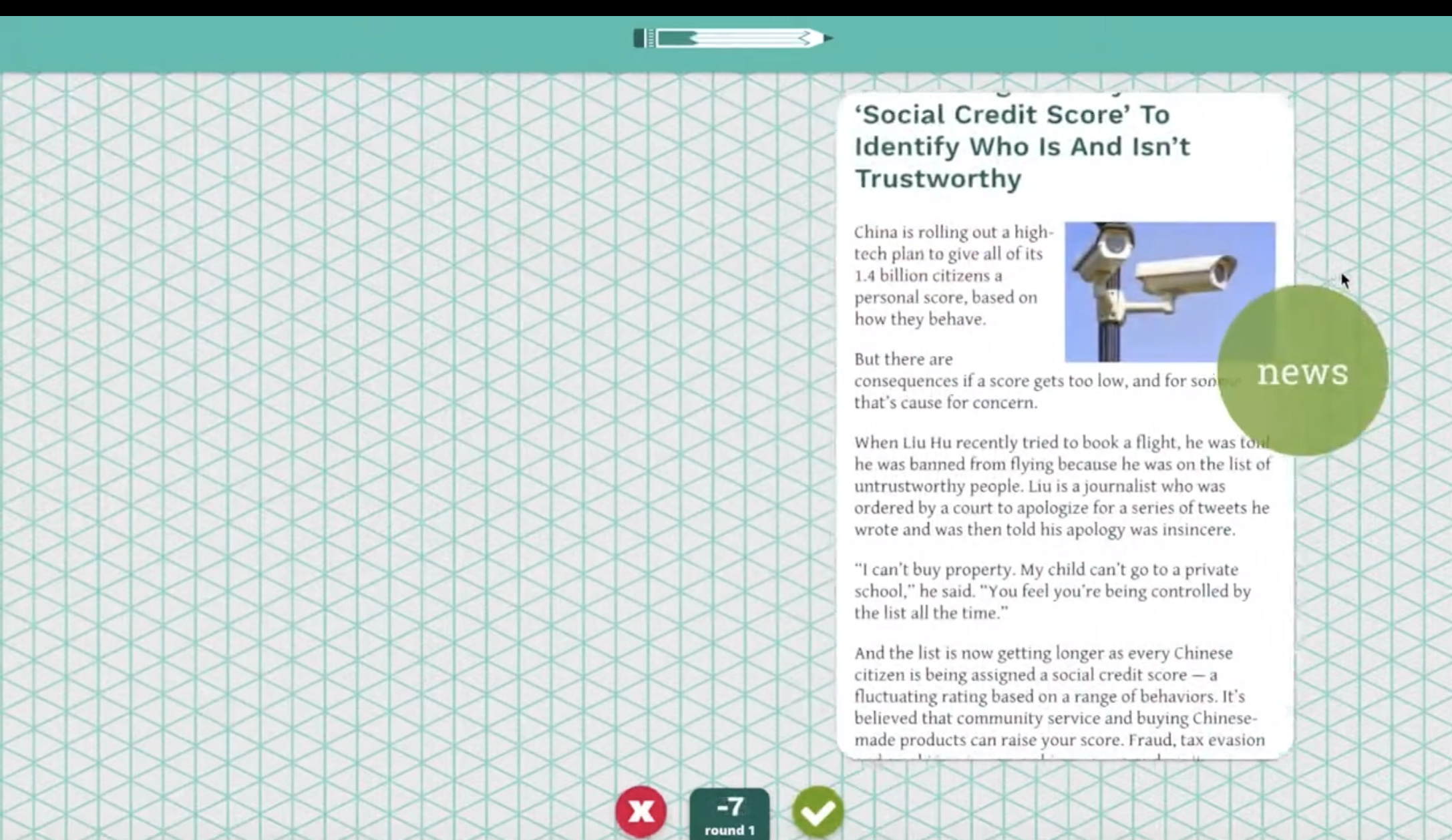}
            \caption{Screenshot from the game \emph{Factitious} (first released in 2017, \cite{grace-2019}), where players discern between real and fake news articles. The game uses real news articles on COVID-19 and health, elections and politics, science and nature, social and cultural issues, satire and clickbait.}
            \Description{Screenshot of the game Factitious displaying a news article headlined "Social Credit Score to identify who is and isn't trustworthy" alongside an image of surveillance cameras and a green "news" label. Along the bottom are a red cross, a round score reading minus seven, and a green check mark for rating the article as fake or real.}
            \label{fig:verificationgame}
        \end{figure*}

We analyze the underlying learning objectives and, where applicable, the theories used in the games. We identified content veracity discernment (misinformation detection) as the most prevalent objective, appearing in $n=31$ (57\%) games (e.g., \emph{Factitious} in Figure~\ref{fig:verificationgame}, \emph{SWRFakeFinder}). These games focus on users’ ability to distinguish between misleading and truthful content, such as posts, headlines, or articles. They frequently employ feedback-driven learning mechanics, aligning with behaviorist principles, and enable procedural knowledge acquisition through learning-by-doing interactions ($n=40$, 74\%).
Awareness of manipulative strategies was present in $n=28$ (52\%) games and was commonly grounded in psychological inoculation theory ($n=26$, 48\%). Inoculation-based designs often position players in the role of misinformation producers to help them understand and recognize manipulative techniques. These games aim to build resistance to manipulation such as sensationalism, trolling, impersonation, and logical fallacies (e.g., \emph{Bad News Game}, \emph{Harmony Square}, \emph{SchlaWiener}, \emph{CrankyUncle}). 

Procedural literacy ($n=20$, 37\%) represented the next most common intervention type (e.g., \emph{LAMBOOZLED!}, \emph{Facts\&Fantasy}). These interventions typically target a broader set of competencies, including structured strategies for source evaluation such as COR \cite{mcgrewLearningEvaluateIntervention2020} (e.g., \emph{Trustme!}), SIFT \cite{caulfield2023verified} (e.g., \emph{Misinformation is Contagious} \cite{barzilai-2025}), and CRAAP \cite{craapBlakeslee2004} (e.g., \emph{How to Spot Fake News}, \emph{Hacking the Research Library} \cite{pun-2017}).
They frequently incorporated declarative knowledge components ($n=16$, 30\%), typically through tutorials or in-game explanations, which were then followed by opportunities for procedural knowledge application.
AI literacy was included in three games.

Additionally, $n=14$ (26\%) games included socio-emotional learning objectives, often implemented through situated learning approaches ($n=12$, 22\%) or narrative-based designs that support socio-emotional processing ($n=4$).

A part of the corpus ($n=12$, 22\%) was categorized as \emph{systems of power}, incorporating situated learning mechanics that illustrate how systems of power influence public discourse through propaganda, news, and institutional actors by casting the player as part of a news or media operation (e.g., \emph{Headliner: Novinews}, \emph{Not for Broadcast}, \emph{The Westport Independent}). A further $n=10$ (18\%) of the games emphasized normative behavior by teaching social norms of online participation, such as responsible sharing.

\subsubsection{Game Designs and Mechanics (RQ3)}
\textbf{Game Formats and Genres.} Trivia games were the most common mechanic ($n=15$, 27\%; players answer questions, discern veracity, or label content, e.g., \emph{Factitious} in Figure~\ref{fig:verificationgame}, \cite{grace-2019}). This was followed closely by text or choice-based simulation $n=12$ (22\%) (players decide between options, often in chat format, with their decision impacting a simulation, often social media statistics, e.g., \emph{Bad News}, Figure~\ref{fig:simulationgamesbadnews} \cite{roozenbeek-2019}) and simulations ($n=8$, 15\%) (players have more agency and can use different game mechanics to influence the simulation, e.g., \emph{Influence Inc.} in Figure~\ref{fig:powergamessystemsinfluenceince}, \emph{The Republica Times}). Other gameplay mechanics, such as simple decision trees ($n=6$, 11\%), point-and-click ($n=3$), and role-playing games (RPGs) ($n=3$), appeared less frequently.

\textbf{Game Mechanics.}
Games in the corpus used multiple game mechanics related to misinformation education, as listed in Figure~\ref{fig:mechanicslabel}, and their co-occurrence with teaching objectives is detailed in Figure~\ref{fig:heatmap}. Most common were game mechanics around evaluating information ($n=23$, 42\%), specifically headlines ($n=13$ (24\%), e.g., \emph{Factitious}), often combined with scoring and feedback mechanics ($n=21$, 39\%).
Social (media) simulations ($n=19$ (25\%), e.g., \emph{Bad News Game}), where player actions affect a simulated social ecosystem, and strong narrative framing ($n=16$ (30\%), e.g., \emph{CatPark}) also appeared often in the corpus.
Further, we observed content creation ($n=15$, 28\%) as a core mechanic, where players created (mis)information themselves (e.g., \emph{Fake It to Make It)}. Sharing decisions ($n=13$, 24\%) or story decisions ($n=11$, 20\%), where players decide if they want to share a piece of content (e.g., \emph{The Republica Times}) or decide between multiple actions to continue the story (e.g., \emph{Choose your own fake news}), often co-occurred with the previous mechanics.
Ten games were competitive, while only three were cooperative (e.g., \emph{Hacking the research library}). Most games were single-player games, sometimes with competitive elements like leaderboards.
While multiple games assigned a role to the player, three put their main focus on role-play (e.g., \emph{Social Media Puppeteers}).

\textbf{Player Role.}
Player positioning strongly influenced both narrative framing and game design, resulting in seven primary player role labels. The most common roles frame the player as a "good", prosocial information citizen ($n=17$ (31\%), e.g., \emph{Fakey}), a detached evaluator ($n=15$ (28\%), e.g., \emph{Factitious}), or an embedded evaluator within the story world ($n=13$ (24\%), e.g., \emph{How to Spot Fake News}). These roles predominantly align players with prosocial and critical-reading perspectives.

\begin{figure*}[h]
            \centering
            \includegraphics[width=0.5\linewidth]{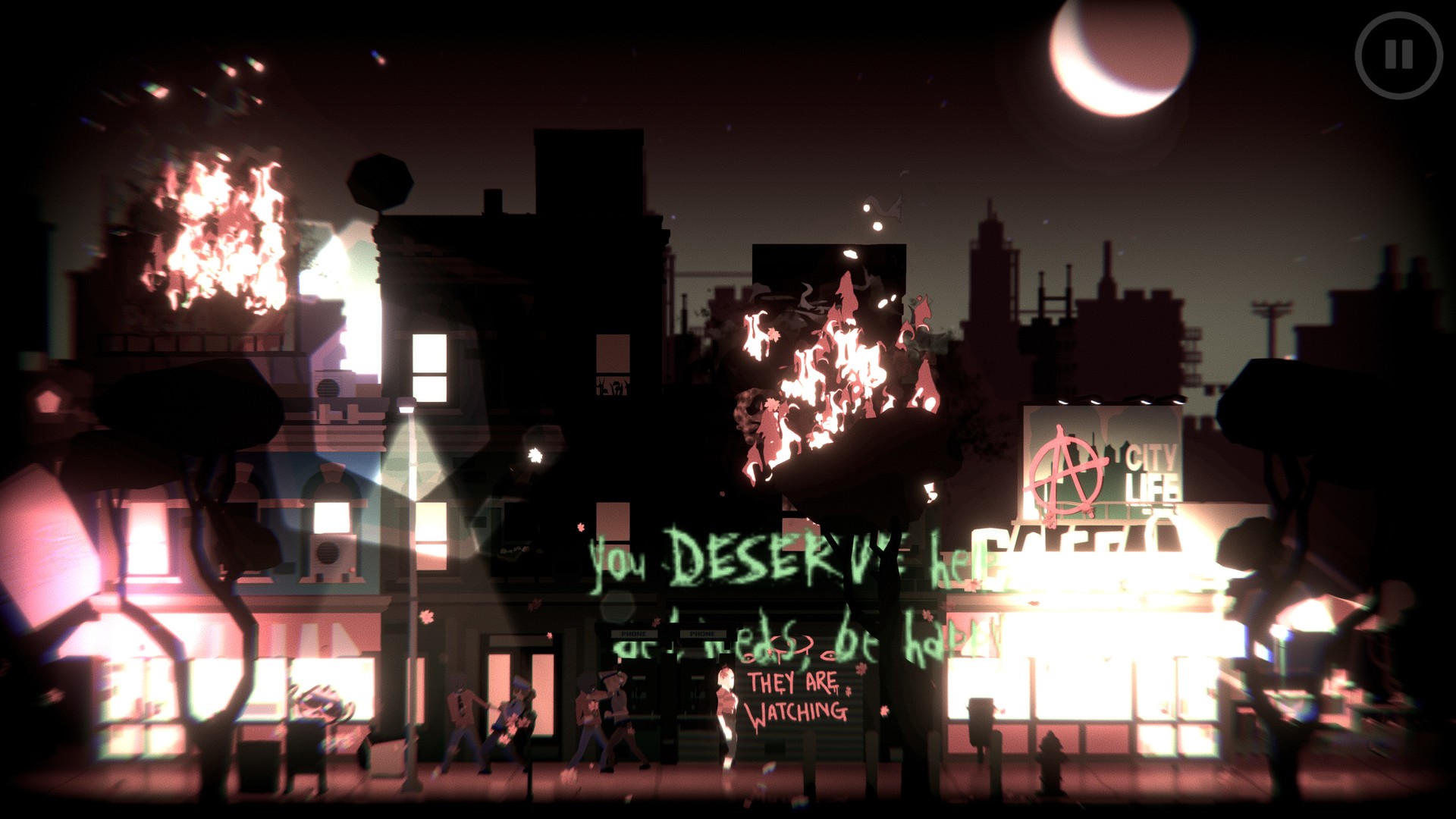}
            \caption{\emph{Headliner:Novinews} (released 2018, © Unbound Creations) depicts the impact of the players’ choices as a national news editor on the city. Players decide which fictional news items and perspectives to publish, ranging from immigration and healthcare to genetic modification, queer rights, alcohol, and vice.}
            \Description{Screenshot from the game Headliner: Novinews showing an atmospheric night-time city street with buildings on fire and protest graffiti reading "You Deserve" and "They are watching", depicting the social fallout of the player's news decisions.}
            \label{fig:powergamessystemsnovinews}
        \end{figure*}

"Good actors" receive the mission to combat misinformation, framing them as morally aligned with this goal. In contrast, "detached evaluators" are tasked with assessing content or making decisions without narrative immersion, often operating as themselves in abstracted or task-oriented environments. "Embedded characters" are more deeply immersed within the narrative context, sometimes without a clearly defined moral alignment, and are required to make decisions as part of the story world, often facing emotional, ethical, or moral consequences (e.g., a city on fire after player decisions in \emph{Headliner:Novinews}, Figure~\ref{fig:powergamessystemsnovinews}).

A subset of games casts players as opponents ($n=11$, 20\%, e.g., \emph{LAMBOOZLED!}, Figure~\ref{fig:cardgamelambozzled}) or bad actors ($n=7$ (13\%), e.g., \emph{Fake It to Make It}, \emph{Bad News Game}). This is particularly evident in content-creation and simulation games, where players actively create or spread misinformation. These designs align with inoculation approaches that place players in manipulative roles to help them understand such strategies.
The distinction between these roles lies in their framing: "opponents" typically compete against other players, often with minimal narrative context, whereas "bad actors" are assigned a fictional role associated with misinformation production. In six opponent-role games (e.g., \emph{FakeYou}), players engage in both creating (misinformation production) and discerning (misinformation detection), although their role engagement remains relatively detached. These games tend to feature limited narrative framing, evoke less emotional involvement, and focus primarily on the mechanics of misinformation creation and detection.

\begin{figure*}[h]
            \centering
            \includegraphics[width=0.7\linewidth]{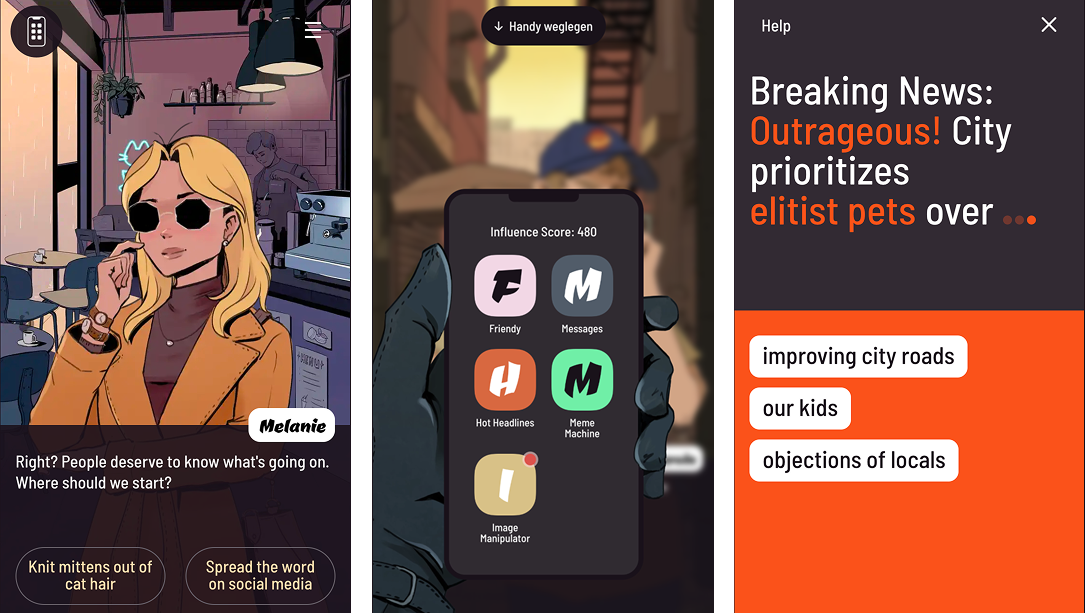}
            \caption{\emph{CatPark} (released 2022 by  the U.S. Department of State's Global Engagement Center, Tilt, Gusmanson, and the University of Cambridge) is a mobile game where players are manipulated into using misinformation strategies to spread the fictional narrative in which the city is controlled by elites and cats. Players themselves create sensational headlines, memes, deepfakes, and manipulated images to push the narrative.}
            \Description{Three screenshots of the mobile game CatPark: Screen one: a stylized blonde character named Melanie in a dialogue "Right? People deserve to know what's going on. Where should we start?" with the answer options "Knit mittens out of cat hair" and "Spread to word on social media", Screen two: a phone interface with app icons, Screen 3: a Headline "Breaking News: Outrageous City prioritizes elitist pets over ... " with three selectable options "improving city roads", "our kids", and "objections of locals".}
            \label{fig:narrativegamecatpark}
        \end{figure*}

Two games (\emph{We Become What We Behold}, \emph{CatPark} in Figure~\ref{fig:narrativegamecatpark}) position players as accidental bad actors, engaging in manipulative strategies unintentionally because of persuasive in-game systems or narrative influences.

\subsubsection{Media Context (RQ4)}
News articles were the most frequently observed media context, appearing in $n=37$ (69\%) games. Of these, $n=21$ (39\%) interventions used news articles exclusively as the media context. 
Social media (e.g., simulated feeds or social media platforms) were present in $n=22$ (41\%) games. A subset of the analyzed games, $n=12$ (22\%), contained both news articles and social media, as their media context.

Moreover, $n=10$ (19\%) games incorporated multimodal formats, defined as including video, images, or audio as their media context. Only $n=5$ games contained videos, and only one had a corresponding paper (SWR FakeFinder \cite{artmann-2023}). The remaining $n=9$ (17\%) games used other media contexts such as conspiracy theories $n=4$, deepfakes $n=3$, or propaganda $n=2$, while $n=2$ games featured no specific media context. One game allowed players to create content with generative AI (\emph{Escape Game}).

\textbf{Types of Misinformation.} $n=42$ (78\%) used non-real, fictional misinformation content such as made-up headlines or posts (Figure~\ref{fig:misinfolabel}). Only $n=15$ (28\%) games used real content from news or social media. Some of the games mixed both, e.g., \emph{FakeYou} used real press images but made-up headlines. Players created the misinformation content themselves in $n=8$ games. The majority of misinformation content was political or politically motivated, even when the content was fictional.

\subsection{Thematic Analysis and Types of Misinformation Games (RQ1)} \label{sec:thematic}
To refine and deepen our analysis of misinformation games for misinformation mitigation (RQ1), we conducted a thematic analysis. Games for misinformation literacy differ in how their mechanics support increasingly sophisticated learning objectives. The identified themes, which are summarized in Table~\ref{tab:misinfo-game-themes}, also reflect a progression in game design complexity. Early themes resemble gamified learning interventions, relying on a small set of mechanics to reinforce discrete verification skills. Later themes increasingly resemble fully fledged games, combining richer narratives, multiple interacting mechanics, player roles, and simulations to support higher-order learning objectives related to the social and systemic dimensions of misinformation. During the thematic analysis, we observed that particular combinations of game mechanics consistently aligned with specific educational objectives \cite{oberdorferGamifiedKnowledgeEncoding2018}, resulting in five recurring game types that progress from claim-level to strategy-level, platform-level, role-level, and ultimately system-level learning.

\begin{table}[h]
\centering
\small
\begin{tabular}{@{}
    p{0.115\textwidth}
    p{0.15\textwidth}
    p{0.13\textwidth}
    p{0.27\textwidth}
    p{0.27\textwidth}
@{}}
\toprule
Theme & Teaching Goal & Theory & Game Mechanics/Roles & Games/Refs \\
\midrule
Verification Games ($n=23$)
 & Claim-level \linebreak detection, source checking
 & Practice-based news literacy drills, inoculation
 & Cue-based quizzes, headline decisions, and real/fake classification with immediate feedback; players act as detached judges or competitors
 & \cite{chang-2020, literat-2021, grace-2019, clever-2020, literat-2020, maekawa-2021, pimmer-2020, katsaounidou-2019, carenzio-2025, pun-2017, buchner-2024, paraschivoiu-2021, artmann-2023, lees-2023}, \emph{Facts \& Fantasy}, \emph{Ezra \& Alex}, \emph{Post Facto}, \emph{Escape Game}, \emph{Real, LOLZ, oops or fake}, \emph{Fight Fakes}, \emph{Exit the Fake}, \emph{Schnitzeljagd: Fake News}, \emph{Choose your own fake news} \\\hline
Strategy Narrative ($n=11$)
 & Teach reusable verification and debunking strategies via stories
 & Narrative-based \linebreak inoculation and strategy-awareness
 & Choice-based dialogues, light RPG progression, mentor/helper characters, narrative framing, gamification of tactics; players act as embedded fact-checkers or bad actors
 & \cite{tandoc-2023, cook-2022, reinhardt-2025, Roozenbeek2019}, \emph{CatPark}, \emph{Actionbound: In the bunker of lies}, \emph{Wiebkes Wirre Welt}, \emph{Facts \& Fakes}, \emph{Digital Matters: Once Upon Online}, \emph{Fake Hunter}, \emph{Fake-News-App: Vorsicht -- Giftstoffe im Handy!} \\\hline
Platform Sims ($n=11$)
 & Show players how posts and shares spread \linebreak misinformation on platforms
 & Situated learning and \linebreak inoculation
 & Newsfeed or social media interfaces with follower/reputation scores and sharing/liking/moderation actions; players are embedded as influencers or moderators in a platform
 & \cite{sureephong-2023, barzilai-2025, micallef-2021, capecchi2024, urban-2018, koutsikou-2025, roozenbeek-2019,basol-2020,scheibenzuber-2019, yang-2020, roozenbeek-2020, ibanez-2024}, \emph{NewsFeed Defenders} \\\hline
Role\linebreak Immersion ($n=3$)
 & Build socio-emotional and misinformation strategy insight by acting out specific roles
 & Role-play and cooperative, situated \linebreak learning
 & Tabletop or digital role-play with alliances, negotiations, and cooperative missions; players take roles such as journalists, influencers, activists, or industry actors
 & \cite{carenzio-2025}, \emph{Doubt is our Product}, \emph{Fake-O-Mat} \\\hline
Systems-Power Games ($n=6$)
 & Reveal propaganda, power, and systemic manipulation
 & Critical media literacy and systems thinking
 & Newspaper or broadcast simulations with management layers and branching political outcomes; players control outlets, regimes, or influence operations that shape public opinion
 & \cite{fabacher-2025}, \emph{We Become What We Behold}, \emph{The Republica Times}, \emph{The Westport Independent}, \emph{Influence Inc.}, \emph{No Place for the Dissident} \\
\bottomrule
\end{tabular}
\caption{Themes of misinformation literacy games by learning objective, underlying learning theory, mechanics, and associated games/citations.}
\label{tab:misinfo-game-themes}
\end{table}

\begin{figure*}[h]
            \centering
            \includegraphics[width=0.8\linewidth]{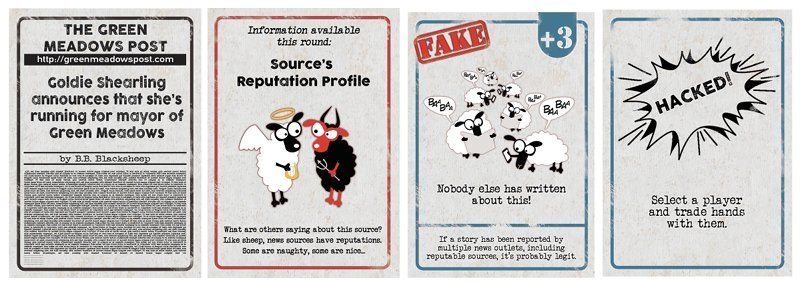}
            \caption{Card Game \emph{LAMBOOZLED!} (released 2020, \cite{chang-2020, literat-2021}). In \emph{LAMBOOZLED!}, players weigh source reputation and identify fake news cards. The game's content is imaginative and focuses on misinformation in the form of sensationalistic and clickbait titles, deceptive links, fake outlets, missing sources, edited or out-of-context images, and misinterpreted data.}
            \Description{Four Cards from the game LAMBOOZLED! are displayed: a fictional newspaper article from The Green Meadows Post, a "Source's Reputation Profile" card showing an angel and devil sheep, a red "Fake +3" card, and a "Hacked! Select a player and trade hands with them" action card.}
            \label{fig:cardgamelambozzled}
        \end{figure*}

\textbf{Verification Games} ($n=23, 42.6\%$) primarily support the learning objective of claim-level verification and source evaluation through simple verification and credibility-scoring mechanics, sometimes in a quiz or trivia format.
The mechanics are intentionally simple because the educational focus is repeated practice in recognizing misinformation cues rather than understanding broader social dynamics.
These designs teach players to judge whether a claim is reliable, often through headline swiping, credibility hints, or repeated real/fake classification with feedback. They tend to have minimal narrative, low game mechanic complexity, and a detached player position, and their main focus is on detection, source evaluation, and veracity discernment. Examples include \emph{Factitious} in Figure \ref{fig:verificationgame}, \emph{SWR FakeFinder Kids} (swipe real/fake), or \emph{Real LOLZ oops or fake} (multi-cue stories).
A subtype of verification themes ($n=9$, 17\%) adds competitive misinformation creation as a game mechanic, where players are opponents discerning each other's misinformation, often in a classroom setting with analogue games (e.g., \emph{FakeYou}, \emph{LAMBOOZLED!} in Figure~\ref{fig:cardgamelambozzled}).

\begin{figure*}[h]
            \centering
            \includegraphics[width=0.7\linewidth]{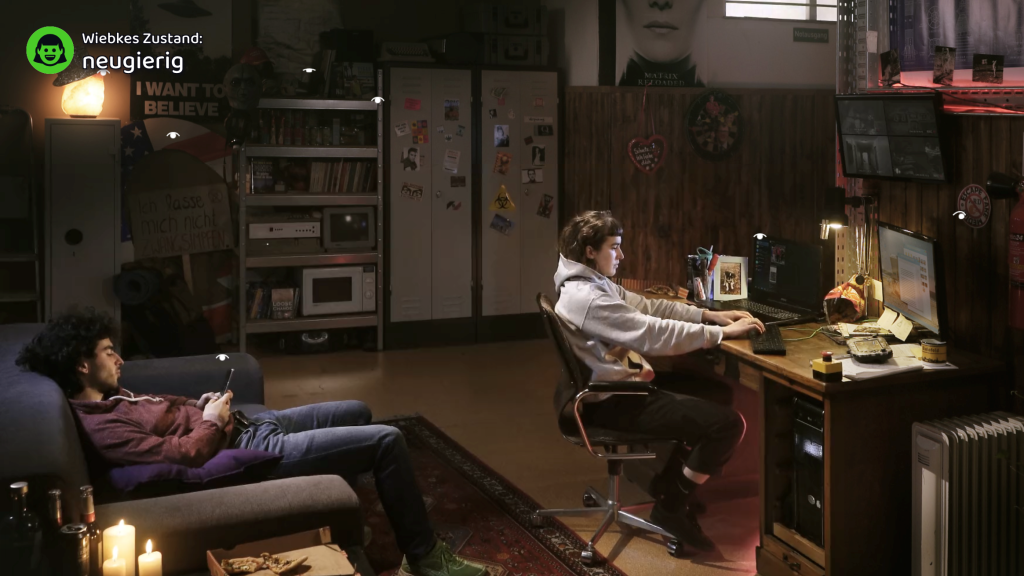}
            \caption{\emph{Wiebkes Wirre Welt} (released 2021, © Kubikfoto GmbH), a point-and-click interactive video in which the player follows a person drawn into conspiracy narratives. Conspiracy topics include secret government plans, the faked moon landing, reptilians, and COVID-19. The game also shows how conspiracy actors downplay politically motivated violence.}
            \Description{Screenshot from the interactive video game Wiebkes Wirre Welt showing a dimly lit teenager's room, with one person lying on a couch and another seated at a desk in front of a computer.}
            \label{fig:narrativegamewiebkeswirrewelt}
        \end{figure*}

\textbf{Strategy Narrative} games ($n=11, 20.4\%$) extend learning beyond isolated fact-checking by teaching reusable verification strategies such as SIFT or CRAAP. Narrative mechanics provide scaffolding and contextual situations in which players practice applying these strategies, making procedural knowledge more transferable than simple verification drills, sometimes with declarative instructions, e.g., for COR.
These games are still relatively simple in structure, but they add emotional and social context so that players can experience misinformation as something with consequences rather than as a purely technical classification task. Examples include \emph{CatPark} in Figure \ref{fig:narrativegamecatpark}, \emph{Fake News Detective} (level-based fact-checker story), \emph{Cranky Uncle} (FLICC denial tactics via uncle mentor), and \emph{Wiebkes Wirre Welt} (Figure~\ref{fig:narrativegamewiebkeswirrewelt}), a point-and-click interactive video on conspiracy narratives.

\textbf{Platform Simulations} ($n=11, 20.4\%$) aim to teach how misinformation spreads through social media ecosystems. Feed-based mechanics expose players to engagement incentives, visibility algorithms, and sharing decisions, helping learners understand diffusion processes.
These games ask them to decide what to share, what to amplify, or how to produce content in feed-like environments, which makes the mechanics closely mirror real platform logics. The focus shifts from merely identifying false claims to understanding how misinformation circulates through social pressure, visibility incentives, and repeated engagement, often with a strong inoculation motive and constructivist, situated learning approach.
Players take on either a bad actor role, where they produce misinformation (e.g., \emph{Bad News} in Figure \ref{fig:simulationgamesbadnews}, \emph{Harmony Square}, exposing tactics like polarization) or a good or embedded actor role (e.g., \emph{Fakey} (like/share/fact-check).

\textbf{Role Immersion} games ($n=3, 5.6\%$) primarily develop socio-emotional and collaborative competencies by placing players in professional or stakeholder roles. Players take on embedded roles such as journalist, influencer, fact-checker, or misinformation producer, and the learning happens through negotiation, discussion, and shared responsibility. This makes the games more narratively and socially complex and often strengthens socio-emotional learning, procedural practice, and situated understanding (e.g., \emph{Social Media Puppeteers} (alliance/rivalry posts), \emph{Doubt is our Product} (industry vs. activists TTRPG)).

\begin{figure*}[h]
            \centering
            \includegraphics[width=0.8\linewidth]{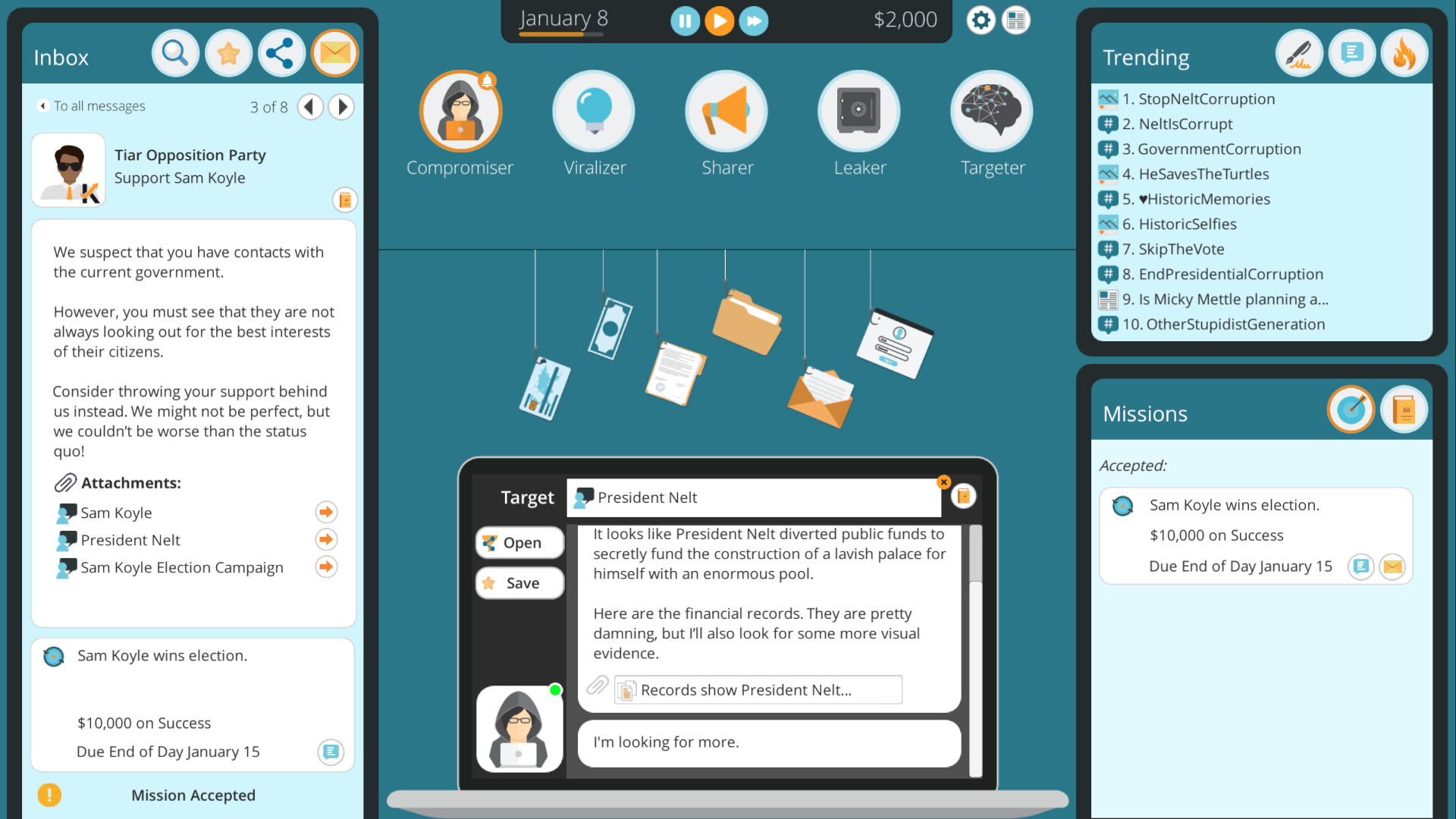}
            \caption{\emph{Influence Inc.} (released 2022, © Curious Bird AB) is a high-fidelity news and social media manipulation simulation in which players act as editors. News topics center on imaginative in-game politics, elections, and war, but also include entertainment, pop culture, business, and technology.}
            \Description{Screenshot of the game Influence Inc. showing a dense dashboard: an inbox with a message prompting the player to support a political candidate, a row of tools labelled Compromiser, Viralizer, Sharer, Leaker and Targeter, a trending list of corruption-related hashtags, a dialogue panel, and a missions panel.}
            \label{fig:powergamessystemsinfluenceince}
        \end{figure*}

\textbf{Systems-Power Games} ($n=6, 11.1\%$) address the highest-level learning objective of understanding mis- and disinformation as a systemic and institutional phenomenon.
These designs focus less on individual checking and more on how misinformation is shaped by institutions, political interests, or platform-wide dynamics. Gameplay and narratives tend to be complex and games are often commercial productions with more high-fidelity mechanics. System-power games frame misinformation as a systemic problem rather than just a problem of false content, and use narrative and simulation to show how manipulation affects society as a whole (e.g., \emph{Headliner Novinews} in Figure \ref{fig:powergamessystemsnovinews}, \emph{Not for Broadcast}, \emph{Influence Inc.} in Figure~\ref{fig:powergamessystemsinfluenceince}).

\subsubsection{Differences of Academic and Non-Academic Games.} \label{sec:results:differences}
Non-academic games tend to place greater emphasis on narrative, embedding players within fictional settings or abstracting complex real-world systems such as media ecosystems or political structures into coherent game worlds. Learning is primarily transported through the narrative and through situated, constructivist experience, with the connection between game mechanics and teaching objective often remaining indirect. Players are more frequently acting as embedded strategic agents, navigating systemic complexity from within rather than receiving explicit instruction or following linear story paths.

Academic games, by contrast, rely heavily on scoring and feedback mechanics, and frequently task players with producing misleading content themselves. By positioning the player in the role of a misinformation producer, the games create a somewhat contradictory relationship between the teaching objective and game mechanic implementation, based on inoculation theory. Players learn how to act as manipulators to understand how misinformation strategies work instead of directly learning how to spot such strategies. However, direct relationships between the teaching objectives and the game mechanics are also more common, reflecting an instructional design orientation more rooted in cognitive and behaviorist teaching approaches. Theoretically, academic games concentrate around inoculation and prebunking, veracity discernment, and bias recognition, with discernment and strategy awareness as the dominant teaching objectives. Players more often take on the role of opponent or detached evaluator, and narrative elements are less prominent or absent, prioritizing procedural skill transfer over immersion and situated, constructivist learning. Figure~\ref{fig:academicvsnonacademic} shows the frequency shifts of labels between academic and non-academic games.

\subsection{Empirical Research on Misinformation Games} \label{sec:academic}
In this section we examine the empirical research conducted on the games. For the 30 games with corresponding publications, we report the contexts in which they were studied, the populations sampled, and the outcomes measured and reported.
\subsubsection{Study Designs and Samples (RQ5)}
We begin with the settings and populations in which these interventions were evaluated. The contextual settings and corresponding study designs of the reviewed papers are summarized in Table \ref{tab:study_contexts}. Within the classroom and formal education context ($n=12$), interventions were deployed in primary, secondary, and university settings. In this context, randomized controlled trials and mixed-methods study designs were used in $n=4$ studies, while within-subjects designs and quasi-experimental comparisons were used in $n=2$ studies. 
Interventions investigated in an online environment ($n=10$) employed a randomized controlled trial design half of the time. The online setting was exclusive to observational field studies ($n=3$). Research situated in a controlled environment, such as a laboratory, used an exploratory user experience and playtesting setup ($n=3$), and interventions were evaluated in workshop settings through participatory design studies. 

\begin{table*}[h]
    \centering
    \small
    \caption{Overview of Study Contexts}
    \label{tab:study_contexts}
    \begin{tabular*}{\textwidth}{@{\extracolsep{\fill}} p{0.3\textwidth} p{0.5\textwidth} r}
        \toprule
        \textbf{Context} & \textbf{Study Design} & \textbf{Total Count} \\
        \midrule
        Classroom / Formal Ed. & Randomized Controlled Trial (RCT) \cite{barzilai-2025, pimmer-2020, reinhardt-2025, roozenbeek-2019} \newline Quasi-Experimental / Causal Comparison \cite{artmann-2023, tandoc-2023} \newline Within-Subjects (Pre-Post) \cite{buchner-2024, koutsikou-2025} \newline Mixed-Methods / Design-Based Research \cite{capecchi2024, cook-2022, maekawa-2021, carenzio-2025} & 12 \\
        \midrule
        Online / Remote & Randomized Controlled Trial (RCT) \cite{basol-2020, lees-2023, roozenbeek-2020, scheibenzuber-2019, yang-2020} \newline Within-Subjects (Pre-Post)/Field Study \cite{sureephong-2023, roozenbeek-2019} \newline Observational Field Studies \cite{grace-2019, katsaounidou-2019, micallef-2021} & 10 \\ 
        \midrule 
        Laboratory (Controlled Setting) & Exploratory UX \& Playtesting \cite{clever-2020, urban-2018, paraschivoiu-2021} & 3 \\
        \midrule
        Workshop & Participatory Design \cite{literat-2020, pun-2017} & 2 \\ 
        \midrule 
        Analytic Review & Qualitative Game Content Analysis \cite{fabacher-2025, ibanez-2024} & 2 \\
        \bottomrule
    \end{tabular*}
\end{table*}

The academic studies predominantly evaluated the interventions in educational settings (see \ref{tab:rq5_results}). The most frequently studied population are K-12 students, with $n=13$ studies using them in interventions ranging from primary to high school. High school students (14--19) are the most frequently researched group. Sample sizes in this context are the largest for middle school cohorts, ranging from 6--217 participants. 
Higher education populations, such as university students, were studied in $n=6$ studies with a population size ranging from 15--84.

\begin{table*}[hb]
    \centering
    \small
    \caption{Overview of Intervention Populations and Demographics}
    \label{tab:rq5_results}
    \begin{tabular}{p{0.25\textwidth} p{0.15\textwidth} c p{0.2\textwidth}}
        \toprule
        \textbf{Population Groups} & \textbf{Demographic} & \textbf{Population Size ($N$)} & \textbf{Reference} \\
        \midrule
        K-12 Students & Primary School\newline (4th Grade) & 29--47 & \cite{artmann-2023, koutsikou-2025} \\
        & Middle School \newline(ages 10--14) & 6--217 &  \cite{chang-2020, literat-2020, capecchi2024, barzilai-2025} \\
        & High School \newline(ages 14--19) & 6--132 & \cite{barzilai-2025, chang-2020, literat-2021, tandoc-2023, Roozenbeek2019, carenzio-2025, paraschivoiu-2021, basol-2020}\\
        \midrule
        Higher Education & University Students \newline(ages 18--30) & 15--84 & \cite{clever-2020, tandoc-2023, pun-2017, cook-2022, buchner-2024, pimmer-2020} \\
        General Public & Adults \newline(ages 18--80+) & 7--681 & \cite{sureephong-2023, urban-2018, roozenbeek-2020, katsaounidou-2019, yang-2020, reinhardt-2025, scheibenzuber-2019} \\
        Large-Scale Online Case Studies & Mixed \newline(ages 10--80+) & 2,847--450,569 & \cite{grace-2019, roozenbeek-2019, lees-2023}\\
        N/A & Qualitative Game Analysis & N/A & \cite{ibanez-2024, fabacher-2025}\\
        \bottomrule
    \end{tabular}
\end{table*}

In contrast to interventions deployed in educational contexts, $n=7$ studies evaluated interventions with the general public or recruited participants via online platforms. These studies are characterized by a wider age range (18--80+) and a larger number of participants (7-681) in their population. However, outliers with small sample sizes ($n=7$) were also observed \cite{urban-2018}.
Large-scale online case studies employ the largest study populations with participant counts ranging from 2,847 to 450,569 \cite{grace-2019, roozenbeek-2019, lees-2023}.

\subsubsection{Metrics and Results (RQ6)}
We now summarize the metrics and results from misinformation game studies. The most frequently reported outcome was veracity discernment, the ability to correctly discern between true and false information. The majority of studies report that their evaluated gamified interventions led to improved discernment skill in participants compared to the respective control groups or baselines \cite{barzilai-2025, grace-2019, Roozenbeek2019, roozenbeek-2019, basol-2020, yang-2020, cook-2022, roozenbeek-2020, artmann-2023, reinhardt-2025, lees-2023}. In some cases, improvements were only observed partially, for example, only in specific topic areas \cite{sureephong-2023, pimmer-2020}.
However, several studies reported no significant improvement regarding their discernment abilities or low learning scores \cite{paraschivoiu-2021, scheibenzuber-2019, buchner-2024}. \citet{barzilai-2025} observed an increase in sharing discernment, resulting from a reduction in willingness to share misleading content while the sharing of reliable information was maintained. One study observed an improvement in detection but simultaneously a decline in the accurate labeling of true news, possibly indicating general skepticism instead of improved discerning abilities \cite{reinhardt-2025}.

Multiple studies reported on the participants' perceived improvement of their skills \cite{clever-2020, pun-2017, buchner-2024, katsaounidou-2019}, although this subjective assessment is not necessarily aligned with objective performance. For instance, \citet{buchner-2024} reported a perceived learning gain but no improvement in discernment ability.

Regarding the User Experience (UX) and Player Experience (PX), qualitative testing reported positive feedback, with minor suggestions for improvements. Participants described the gamified interventions positively  \cite{cook-2022, chang-2020, micallef-2021}. Quantitative assessments using the System Usability Scale (SUS) were mostly favorable \cite{koutsikou-2025} or slightly below the optimal threshold \cite{capecchi2024}. Furthermore, when the interventions were compared to text-based approaches, participants expressed a preference for the gamified interventions \cite{reinhardt-2025, scheibenzuber-2019, katsaounidou-2019}.

\section{Discussion}
This systematic mapping review presents, to the best of our knowledge, the largest corpus of misinformation literacy games analyzed to date, comprising 54 games. The dataset supplements 30 academically documented games with 24 non-academic games, thereby broadening the analytical lens beyond peer-reviewed literature. The overall increase in identified games compared to prior reviews supports the hypothesis that the field is expanding \cite{Rev1_2024_Tackling}. In the following, we first discuss patterns and trends in game designs, then focus on academic evaluation practices, and conclude with identifying structural gaps and opportunities for future research and game development.

\subsection{Patterns of Misinformation Games}
\textbf{Free-To-Play Browser Games and Classroom Interventions (RQ1).}
Regarding types of misinformation games, the majority are digital video games, predominantly web-based, free to play, easily accessible via browsers, but with rather simple game mechanics. Only a small number, such as commercial titles like \emph{Influence Inc.}, have more complex game designs and require purchase. Notably, some digital games are browser-accessible on mobile but not optimized for it, reducing (touch) usability and ecological validity. As misinformation consumption occurs primarily on smartphones via social media feeds, mobile-first design would better match the context these games aim to simulate.
Non-digital games are typically designed for classroom settings. These interventions often integrate directly into formal education contexts, suggesting a clearer pedagogical embedding but potentially limiting broader dissemination.

\textbf{Focus on Narrow Learning Objectives (RQ2).}
In terms of learning objectives and theoretical backgrounds, most games target general detection and awareness of misinformation and manipulation strategies as their primary learning objective. Through learning approaches based on behaviorism and cognitivism, they focus narrowly on a limited subset of competencies. Core skills such as source evaluation (\cite{McGrew-2018}), lateral reading \cite{Wineburg-2019}, cognitive reflection \cite{pennycook-2019}, algorithmic awareness, and deeper social media literacy \cite{kozyreva-2020} are underrepresented. This is in line with earlier observations of a "sourcing gap" in games' learning objectives \cite{Rev1_2024_Tackling}.
The dominance of inoculation-theory-based \emph{Platform Simulations}, possibly inspired by \emph{Bad News}, and simple \emph{Verification and Strategy Narrative Games} shows a clear theoretical convergence. While inoculation frameworks provide a strong and empirically supported foundation, the lack of theoretical diversity may limit innovation in learning design, reach, and impact of games.

\textbf{Simple Game Mechanics and Low Complexity (RQ3).}
After analyzing and thematically grouping games on their designs and mechanics, we find a strong prevalence of trivia-style and discernment-based \emph{Verification Games}, especially in academic games (see Section \ref{sec:results:differences}). The predominant theme (42.6\%) consists of repeated veracity judgments or gamified quiz structures. These designs align with behaviorist learning paradigms based on repetition and reinforcement. While such approaches are comparatively easy to implement and scale, they may limit player agency and reduce opportunities for deeper cognitive engagement \cite{Rev4_2023_PlayingFakeState,GreitzerCognitiveScienceForGamification2007, arnabMappingLearningGame2015}.
More cognitively demanding designs, such as \emph{Platform Simulations} or \emph{Strategy Narratives}, are present but rarer, e.g., \emph{Bad News} \cite{roozenbeek-2019}. However, they often provide limited agency within predefined decision trees. More open-ended simulations, sandbox-style interactions, or narrative-driven role-play could enhance learning and player ownership of decisions \cite{GeeGamesLearningLiteracy2003}.
Narratively, academic interventions often do use no narratives or narratives of simple to moderate complexity. Where narratives are present, they are typically linear and offer limited replayability.
Player roles further reflect limited agency. Most academic games position players as detached evaluators, good actors, or bad actors, while non-academic games embed players in the setting without predefined paths, requiring moral judgment rather than correct-answer selection.
Few games allow players to define their own moral stance or strategic approach. Increasing player autonomy could enhance emotional investment and learning outcomes by fostering stronger identification and self-determination \cite{krathRevealingTheoreticalBasis2021}. However, more complex game designs come with higher developmental costs, which are often difficult to accommodate in academic research settings.

\textbf{Focus on News Posts (RQ4).}
Games are similar in the media context and content they cover. 
Most center around imaginative or player-created news articles or news-like social media posts on politics, health, or public events. Approximately half of the games present content explicitly in social media formats, reflecting current media consumption patterns. A small subset of games include real posts or news articles. However, only five games incorporate video content, with merely one having a corresponding paper. This represents a substantial gap, given the increasing dominance of misleading short-form video content on platforms such as TikTok, Instagram, and YouTube \cite{GrosseKampmann2025, Jiang2024}. The limited integration of videos may reflect greater developmental complexity, yet it leaves a critical dimension of misinformation literacy underexplored. Similarly, generative AI and deepfakes are addressed in only three games. As AI-generated images and videos increase \cite{lopez2025}, this gap suggests a need for updated educational and design priorities.

\subsection{Scientific Testing of Misinformation Games}
\textbf{Samples and Target Groups (RQ5).}
Compared to earlier reviews \cite{Rev1_2024_Tackling}, our corpus shows a relatively balanced distribution across age groups and research contexts (online versus classroom-based studies). Many games tested with adult samples may also be suitable for adolescents. However, primary and middle school populations remain underrepresented, despite increasing early smartphone and social media use among younger children \cite{Ofcom2024}. Preventive literacy interventions arguably need to begin earlier. The age gap is mirrored in platform selection. Youth-centered ecosystems, such as Roblox, are not represented. However, embedding media literacy education within the social platforms and contexts where misinformation is actually encountered might increase their educational impact.

\textbf{Measuring Veracity Discernment and User Experience (RQ6).}
Most studies measuring veracity discernment report improvements following gameplay, suggesting that misinformation games are generally effective interventions. However, measurement heterogeneity limits comparability, as widely varying metrics for veracity assessment are used. Greater standardization, potentially via instruments such as the MIST-20 \cite{maertens-2023}, could increase comparability and knowledge building. At the same time, the field must critically reflect on whether headline-based veracity test tasks adequately capture multimodal media competencies, particularly regarding videos, feeds, and generated content.
Some studies rely solely on self-reported learning gains. One study \cite{buchner-2024} reports discrepancies between perceived improvement and objective ability improvement, questioning the validity of self-report measures as standalone metrics.

UX and PX measures are included in some studies, yet only three games combine objective veracity assessment with UX/PX measures. We argue that educational effectiveness and UX/PX should not be treated as separate concerns. Games that are impactful but not engaging will fail to scale; conversely, enjoyable games without measurable impact miss the educational objective. We advocate for dual evaluations integrating both learning metrics and UX/PX metrics.

\subsection{Academic vs. Non-Academic Games.}
Academically developed games are more frequently evaluated for learning effects, sometimes with multiple studies. However, many are no longer publicly available or never reach broader audiences. If games remain confined to experimental settings, their real-world impact remains limited.
Non-academic productions, by contrast, sometimes benefit from professional production teams which may increase aesthetic quality, playtime, and replayability. These games may achieve higher hedonic standards but often lack rigorous evaluation or more direct implementation of learning objectives.
Therefore, we argue for stronger collaboration between academia, educators and professional game development. Scientifically grounded interventions combined with high production quality and effective dissemination strategies might lead to a higher impact and a broader reach.

\subsection{Limitations}
The findings presented in this review are limited by the following factors. First, the search strategy relied on Google Scholar for academic literature, possibly missing entries from other platforms. The reproducibility of the qualitative Google search results for non-academic games are limited due to the location-dependent nature of the search algorithm. Therefore, relevant games may have been missed due to algorithmic bias. Furthermore, while coding was conducted systematically, coder error or individual differences influencing interpretation remain potential sources of bias.
Additionally, the review was limited by the availability of games investigated in academic studies. If games were unavailable, they were coded based on secondary information and otherwise excluded.

\subsection{Gaps and Future Work}
Our corpus showed several gaps in the realm of misinformation games:
\begin{enumerate}
    \item Video-based misinformation as a topic is substantially underrepresented in the corpus.
    \item Generative AI as a subtopic of misinformation is nearly absent in the corpus.
    \item Emotional and narrative depth of games in the corpus is limited but might increase learning effects.
    \item Player agency and systemic simulations remain rare across analyzed games but might increase learning effects.
    \item The heterogeneity of measurements in corresponding studies limits the comparability of games.
    \item The reporting quality of papers describing games varies, with often insufficient documentation of mechanics and games being unavailable.
\end{enumerate}

The field currently shows an overconcentration of relatively simple discernment games and guided text-based simulations. While these formats are valuable and successful \cite{grace-2019, roozenbeek-2019}, greater diversity in mechanics, narrative structures, and experiential depth might be necessary.
On the methodological side, standardized and multimodal assessment frameworks are needed. Additionally, UX/PX testing should be incorporated into evaluation study designs.

\section{Conclusion}
In this systematic mapping review, we analyzed a corpus of 54 gamified interventions designed to counter misinformation. We provide a thematic categorization based on their mechanics, theoretical foundations, player roles, and learning objectives, as well as evaluation findings for the subset of academically evaluated games. 
Our analysis shows a dominance of web-based games, primarily trivia and choice-based simulation formats. Regarding media context, textual news is most commonly used, indicating a strong focus on text-based discernment tasks rather than the multimodal complexity of modern social media environments. 
Reported evaluation outcomes suggest that while games can improve discernment accuracy, general skepticism may also arise as a confounding variable. Nevertheless, evaluated interventions show encouraging effectiveness. Future gamified interventions should focus more on users' contemporary needs and diversify game design beyond text, incorporating multimodal environments that better reflect the visual, auditory, and social complexity of modern digital ecosystems.


\begin{acks}
This work was funded by the ``Daimler und Benz Stiftung'' grant: Ladenburger Kolleg (Project: KonCheck). We acknowledge support by the Open Access Publication Fund of Rhine-Waal University of Applied Sciences. Further, this work was partially supported by JSPS KAKENHI Grant JP24H00732, by JST CREST Grants JPMJCR20D3 and JPMJCR2562 including AIP challenge program, and by JST K Program Grant JPMJKP24C2 Japan. It was also funded by the German Federal Ministry of Education and Research (BMBF) through the DAAD (German Academic Exchange Service).
\end{acks}

\bibliographystyle{ACM-Reference-Format}
\bibliography{sample-base}

\appendix
\onecolumn

\section{Overview of Analyzed Academic Games}

{\footnotesize
\begin{longtable}{
    >{\RaggedRight\arraybackslash}p{0.015\textwidth}
    >{\RaggedRight\arraybackslash}p{0.03\textwidth}
    >{\RaggedRight\arraybackslash}p{0.13\textwidth}
    >{\RaggedRight\arraybackslash}p{0.15\textwidth}
    >{\RaggedRight\arraybackslash}p{0.21\textwidth}
    >{\RaggedRight\arraybackslash}p{0.35\textwidth}
    >{\RaggedRight\arraybackslash}p{0.03\textwidth}
}

\caption{Reviewed papers}
\label{tab:rev_papers} \\

\toprule
\textbf{ID} & \textbf{Year} & \textbf{Game} & \textbf{Study Design} &
\textbf{Population Demographics} & \textbf{Reported Key Findings} & \textbf{Ref} \\
\midrule
\endfirsthead

\toprule
\textbf{ID} & \textbf{Year} & \textbf{Game} & \textbf{Study Design} &
\textbf{Population Demographics} & \textbf{Reported Key Findings} & \textbf{Ref} \\
\midrule
\endhead

\bottomrule
\endlastfoot

        1 & 2023 & How to Spot Fake News & Within-Subject online study (pre-/post-test) & $n=351$ ages 20-49 & Improvement of overall knowledge on information literacy, spotting fake news in the context of education, health, and religion. No improvement in the context of economic and social news & \cite{sureephong-2023} \\ 
        2 & 2025 & Misinformation is Contagious & Between-Subjects online study & $n=132$ 8th and 9th graders (mean age of 14.23) in Israel & Simulating evaluation strategies improved post-game accuracy and sharing discernment, In-game sharing accuracy mediated post-game performance, misinformation explanations alone did not improve post-intervention performance & \cite{barzilai-2025}\\
        3 & 2021 & LAMBOOZLED! & Iterative design with playtesting sessions and interviews with educators & Playtest $n=48$ sixth graders at an all-boys private school, $n=28$ high school students grades 9-12, interviews with $n=11$ educators & Students successfully transferred news literacy skills to real-life contexts, majority of students perceived the game as valuable and engaging to play. Playtest findings confirmed by educator interviews. Teachers reported that the fictional setting of the game increased engagement and transfer. Educators degree of preparation and perception towards games as learning tools were identified as success factors & \cite{chang-2020, literat-2021} \\
        4 & 2019 & Factitious & Cross-Sectional Analysis (collected via the game) & $n=450,469$ unique players, $n=45.031$ provided demographics, age 10-80+, $n=22,490$ male, $n=21,329$ female, $n=1,212$ other/NB, $n=20,780$ Bachelors , $n=11,556$ Masters, $n=4279$ PhD, $n=8,416$ other & Performance in identifying fake news improved with age (up to age 70); Higher education levels correlated with better performance scores, younger player ages 10-19 made decisions significantly faster than older groups & \cite{grace-2019} \\ 
        5 & 2020 & FakeYou & Multiplayer Playtest with a post survey & $n=53$ university affiliated participants, 75\% male & Small self-reported improvement in figuring out correct headlines, most users had fun, suggested improvements in bugfixes, server performance & \cite{clever-2020} \\
        6 & 2020 & Fakeopoly & Participatory Design Workshop \footnote{Redesign of the board-game Monopoly into a News Literacy Game} & $n=6$ ages 10-14 & Focuses on the participatory game design as a research method, the games were not evaluated.& \cite{literat-2020}\\
        7 & 2020 & The Lying Geese \footnote{Redesign of the card-game Buffalo into a News Literacy Game} & Participatory Design Workshop &$n=6$ ages 10-14 & ocuses on the participatory game design as a research method, the games were not evaluated. & \cite{literat-2020}\\
        8 & 2020 & Brain Company Inspired (Social Media Simulator) & Preliminary Survey and iterative prototyping with a User tests & $n=566$ ages 15-24 (survey), $n=13$ first year university students (user test), $n=23$ students (second user test) & The game triggered self-reflection on social media behavior and change their perception towards evaluation criteria & \cite{maekawa-2021} \\
        9 & 2020 & SWR FakeFinder & Randomized Controlled Trial & $n=72$ bachelor degree students & significant but small increase in news classification skills compared to the pre-test& \cite{pimmer-2020} \\
        10 & 2023 & Fake News Detective & Between-Subjects and Within-Subjects (two studies) & $n=84$ College Students, $n=60$ Secondary and Junior High School Students in Singapore & Treatment group improved self-reported scores on threat perception, skepticism and caution in the context of online information, treatment group scored higher in fake news detection compared to control group, results were replicated across both college and secondary school samples & \cite{tandoc-2023} \\
        11 & 2021 & Fakey & Google Analytics data collection for 19 months and Interviews & $n=6$ Interview Participants 19–46 years old & \cite{micallef-2021} \\
        12 & 2024 & Social4School & & $n=217$ Middle School Students ages 12-14 in Northern Italy& Usability was rated 66.7/100, slightly below the optimal threshold; Suitable educational tool in classrooms with high engagement; Participants reported positive knowledge improvement & \cite{capecchi2024} \\
        13 & 2018 & Fake It to Make It & Interviews and Retrospective think-alouds & $n=7$ & Results show great variation in attitude towards the game &\cite{urban-2018} \\
        14 & 2025 & Fake News Travel Fast & Within-Subjects study & $n=47$ Elementary School Students & Participants showed significant gains in their critical evaluation and skepticism towards online information. Good Usability with a SUS of 75.31/100 & \cite{koutsikou-2025}\\
        15 & 2019 & The Fake News Game & Randomized field study (post-test only) & $n=95$ high school students ages 16-19 in the Netherlands & Treatment group rated reliability, of fake news articles in the context of the refugee crisis, significantly lower than the control group, persuasiveness was rated lower in the treatment group but not statistically significant & \cite{Roozenbeek2019}\\
        16 & 2019 - 2020 & Bad News Game & Automatic user data collection for 3 months & 14,163 - 14,266 online users & Gameplay improved ability to spot and resist misinformation techniques, effects were robust against age, gender and political ideology & \cite{roozenbeek-2019}\\
        & & Bad News Game & Mixed-Method Design, 2 (Bad News/ Tetris) x 2 (Pre-/Post-Survey) & $n=196$, modal age 18-24, 58\% male  & Improvement in spotting misinformation techniques compared to the control group (Tetris), increased confidence in reliability judgement & \cite{basol-2020} \\
        & & Bad News Game & Between-Subjects study & $n=71$ German/Swiss adults avg. age 29 & No significant difference in objective knowledge gain compared to control group (reading an informational text) ; perceived learning effect was higher for the game group & \cite{scheibenzuber-2019} \\ 
        17 & 2019 & MAthE & Online field study (pre-/post-game questionnaire) & $n=111$ participants in the online survey, $n=35$ participants in class discussions & Participants report that they prefer gamified expierences over other learning formats, 66.4\% agree that the game contributes to threir news-verification training, 70.9\% learned about the existence of verification tools, 50.9\% report that it helps them to learn the usage of such tools, 52.8\% report that the game supported them to identifiy fake images and 39.1\% understand the trails of visual manipulation & \cite{katsaounidou-2019}\\
        18 & 2018 & Trustme! & Between-Subject study (Game/Quiz/Control) & $n=210$ adolescents ages 20-29 in South Korea  & Enhanced information discernment skills; Did not improve skepticism toward online information & \cite{yang-2020}\\
        19 & 2025 & Social Media Puppeteers & Qualitative testing & $n=30$ students ages 16-19 & Findings focus on gameplay mechanics and improvement but do not further discuss the efficacy of the intervention & \cite{carenzio-2025} \\
        20 & 2025 & Data Defenders & Qualitative testing & $n=6$ ages 15-18 & Findings focus on gameplay mechanics and improvement but do not further discuss the efficacy of the intervention & \cite{carenzio-2025}\\ 
        21 & 2025 & Social Media Fake News & Qualitative testing & $n=16$ ages 16-19, $n=19$ ages 18-20, $n=12$ ages 17-19 & Findings focus on gameplay mechanics and improvement but do not further discuss the efficacy of the intervention & \cite{carenzio-2025}\\
        22 & 2017 & Hacking the research library & Workshops & $n\approx15$ university students per Workshop & Students felt more confident conducting library research, they reported after the intervention to spend more time verifying sources in news articles in order to assess credibility, better understanding of the possibility of manipulation on wikepedia pages. Although it was observed that students struggled discerning between real and fake reports & \cite{pun-2017} \\
        23 & 2021 & Cranky Uncle & Exploratory study analyzing classroom implementation (Qualitative Caste Study) & Case Study 1: undergraduate course at a Community College in Massachusetts with six classes of $n\approx20$ students with an average age of 22 , Case Study 2: A\&M University Texas class typically attended by $n\approx50-60$ with an average age of 21, Case Study 3: N/A & Case Study 1: increased student engagement and helped students identify misleading arguments, Case Study 2: Class feedback suggests successful use of critical thinking to identify misinformation, Case Study 3: positive feedback from participants & \cite{cook-2022} \\
        24 & 2020 & Harmony Square & Mixed-Randomized control trial & $n=681$ International participants 41,4\% 18-24 years old, 62.4\% with at least a Bachelor's degree & Gameplay reduced the perceived reliability of manipulative content; Increased confidence in spotting manipulation, reduced willingness to share such content & \cite{roozenbeek-2020}\\
        25 & 2024 & Escape Fake & Within-Subjects Design (Pre-/Post-test design with one group) & $n=45$ pre-service teachers with a mean age of 22.59 years & Increased participants knowledge about fake news, more critical towards online information. Furthermore, increased confidence in their own ability to identify fake news, while no significant improvement in discernment skills could be observed  & \cite{buchner-2024} \\
         & 2021 & Escape Fake & User Experience study & $n=49$ adolescents ages 14-18 & Participants positively and reported a high willingness to play the game at home for entertainment. Self-reported rating for learning effectiveness were low & \cite{paraschivoiu-2021} \\
        26 & 2023 & SWR FakeFinder Kids & Mixed Methods design & $n=29$ fourth grade elementary school students in Germany & Significant improvemtent in pre-post test performance for both groups. Cognitive processing in the context of credibility assessment shifted from intuitive to analytical processing & \cite{artmann-2023} \\
        27 & 2025 & SchlaWiener & Randomized Controlled Trial & $n=373$ ages 15-25 & Perceived learning enjoyment was higher for the game-group (active inoculation) compared educational leaflet group (passive inoculation). Participants from both groups improved significantly in detecting fake news, while at the same time a decline in real news detection was observed. & \cite{reinhardt-2025} \\
        28 & 2024 & Headliner: Novinews & Qualitative analysis of the game & N/A &  & \cite{ibanez-2024} \\ 
        29 & 2025 & Not for Broadcast & Qualitative analysis of the game & N/A & & \cite{fabacher-2025} \\
        30 & 2023 & Spot The Troll & Experimental Study, Observational Study & $n=2847$ (experimental study) $n=505$ (observational study) & Significantly increased accuracy in identifying inauthentic (troll) accounts. Despite lower reported self-efficacy, own accuracy was estimated higher compared to control group. Intervention lead to a decrease in sharing behavior.  & \cite{lees-2023} \\
\end{longtable}
}


\section{Overview of All Analyzed Games}

{\footnotesize
\begin{longtable}{
    >{\RaggedRight\arraybackslash}p{0.12\textwidth}
    >{\RaggedRight\arraybackslash}p{0.21\textwidth}
    >{\RaggedRight\arraybackslash}p{0.12\textwidth}
    >{\RaggedRight\arraybackslash}p{0.1\textwidth}
    >{\RaggedRight\arraybackslash}p{0.14\textwidth}
    >{\RaggedRight\arraybackslash}p{0.15\textwidth}
    >{\RaggedRight\arraybackslash}p{0.1\textwidth}
}

\caption{Game types, designs, formats and Media-Context}
\label{tab:RQ1,RQ3,RQ4}  \\

\toprule
\textbf{Game}  & \textbf{Topic} & \textbf{Platform and Intervention Type} & \textbf{Game Type} & \textbf{ Media Context} & \textbf{Player Role} & \textbf{Prior Reviews} \\
\midrule
\endfirsthead

\toprule
\textbf{Game} & \textbf{Topic} & \textbf{Platform and Intervention Type} & \textbf{Game Type} & \textbf{ Media Context} & \textbf{Player Role} & \textbf{Prior Reviews}\\
\midrule
\endhead

\bottomrule
\endlastfoot
        
        How to Spot Fake News &  (Mis)Information Literacy, Social Media & Video Game & Choice-based Simulation & News Articles, Social Media & Good Actor & -- \\
        Misinformation is Contagious &  Social Media, (Mis)Information Literacy & Video Game & Choice-based Simulation & Social Media & Good Actor & -- \\
        LAMBOOZLED! &  (Mis)Information Literacy& Card Game & Card Game & News Articles, Other & Detached Evaluator & -- \\
        Factitious &  (Mis)Information Literacy & Video Game & Trivia & News Article & Detached Evaluator & \cite{Rev1_2024_Tackling, Rev4_2023_PlayingFakeState, Rev3_2024_LiteracyAtPlay} \\
        FakeYou &  (Mis)Information Literacy& Video Game & Trivia &  News Articles, Multimodal (e.g., Video, Image, Audio) & Detached Evaluator & \cite{Rev1_2024_Tackling}\\
        Fakeopoly &  (Mis)Information Literacy& Board Game & Board Game & News Article & Bad Actor & -- \\
        The Lying Geese & (Mis)Information Literacy & Card Game & Card Game & News Article & Strategic Actor & --\\
        Digital BrainCompany Inspired &  (Mis)Information Literacy& Video Game & Trivia & News Article & Strategic Actor & -- \\
        SWR FakeFinder &  (Mis)Information Literacy, Trolls, Satire, Humor, Ads& Video Game & Trivia & News Article & Detached Evaluator & \cite{Rev1_2024_Tackling}\\
        Fake News Detective &  (Mis)Information Literacy & Video Game & RPG & News Article & Good Actor & --\\
        Fakey &  (Mis)Information Literacy& Video Game & Trivia & News Article & Detached Evaluator & \cite{Rev3_2024_LiteracyAtPlay, Rev2_2023_EvaluatingAsTools, Rev4_2023_PlayingFakeState}\\
        Social4School & (Mis)Information Literacy & Video Game & Trivia & News Article & Detached Evaluator & -- \\
        Fake It to Make It &  (Mis)Information Literacy& Video Game & Choice-based Simulation & News Article & Bad Actor & \cite{Rev3_2024_LiteracyAtPlay, Rev2_2023_EvaluatingAsTools, Rev4_2023_PlayingFakeState} \\
        Fake News Travel Fast & (Mis)Information Literacy & Video Game & Choice-based Simulation & Other & Detached Evaluator & -- \\
        The Fake News Game &  Manipulation Strategies, (Mis)Information Literacy & Card Game & Card Game & News Article & Bad Actor & \cite{Rev1_2024_Tackling, Rev3_2024_LiteracyAtPlay, Rev3_2024_LiteracyAtPlay, Rev4_2023_PlayingFakeState}\\
        Bad News Game &  (Mis)Information Literacy, Manipulation Strategies, Trolls, Satire, Humor, Ads& Video Game & Choice-based Simulation & News Articles, Social Media & Bad Actor & \cite{Rev1_2024_Tackling, Rev3_2024_LiteracyAtPlay, Rev2_2023_EvaluatingAsTools, Rev4_2023_PlayingFakeState} \\
        MAthE & (Mis)Information Literacy & Video Game & Other & News Article & Detached Evaluator & \cite{Rev1_2024_Tackling} \\
        Trustme! &  (Mis)Information Literacy& Video Game & Choice-based Simulation, Trivia & Social Media & Good Actor & \cite{Rev1_2024_Tackling} \\
        Social Media Puppeteers &  Social Media & Tabletop Roleplay Game & RPG, Board Game & Social Media & Strategic Actor & -- \\
        Data Defenders &  (Mis)Information Literacy & Video Game & Tower Defense & Social Media, Other & Good Actor & \cite{Rev3_2024_LiteracyAtPlay}\\
        Social Media Fake News &  Manipulation Strategies, (Mis)Information Literacy, Social Media & Card Game & Card Game & News Articles & Bad Actor & --\\
        Hacking the research library &  (Mis)Information Literacy& Escape Room & Other & News Articles, Social Media, Other & Good Actor & --\\
        Cranky Uncle &  (Mis)Information Literacy& Video Game & Decision Tree & No Specific Media Context & Good Actor & \cite{Rev2_2023_EvaluatingAsTools, Rev3_2024_LiteracyAtPlay}\\
        Harmony Square &  (Mis)Information Literacy, Manipulation Strategies, Trolls, Satire, Humor, Ads, Social Media, Bots & Video Game & Choice-based Simulation & News Articles, Social Media & Bad Actor & \cite{Rev4_2023_PlayingFakeState, Rev2_2023_EvaluatingAsTools, Rev3_2024_LiteracyAtPlay, Rev1_2024_Tackling} \\
        Escape Fake &  Bots, (Mis)Information Literacy, Social Media & AR Game & Other & News Articles, Social Media, Multimodal (e.g., Video, Image, Audio) & Good Actor & \cite{Rev3_2024_LiteracyAtPlay, Rev1_2024_Tackling}\\
        Facts \& Fantasy & (Mis)Information Literacy, Conspiracy Theories, Disinformation & Card Game & Card Game & Other & Bad Actor, Good Actor & -- \\ 
        Actionbound: In the bunker of lies &  Social Media, Conspiracy Theories, (Mis)Information Literacy, Manipulation Strategies & Video Game & Choice-based Simulation, Trivia & News Articles, Social Media & Good Actor & -- \\
        Wiebkes Wirre Welt &  (Mis)Information Literacy, Conspiracy Theories, Manipulation Strategies, Social Media & Video Game & Point and Click, Decision Tree & Social Media, News Articles, Multimodal (e.g., Video, Image, Audio) & Bad Actor & --\\
        Facts \& Fakes &  (Mis)Information Literacy& Video Game & Choice-based Simulation & Multimodal (e.g., Video, Image, Audio), Social Media, News Articles & Good Actor & -- \\ 
        CatPark Game & (Mis)Information Literacy, Conspiracy Theories, Social Media, Manipulation Strategies & Video Game & Decision Tree & News Articles, Social Media & Bad Actor & \cite{Rev2_2023_EvaluatingAsTools} \\
        We Become What We Behold &  Polarization & Video Game & Other, Point and Click & No specific Media Context & Bad Actor & --\\
        Digital Matters: Once Upon Online &  (Mis)Information Literacy & Video Game &Decision Tree & News Articles & Detached Evaluator & -- \\
        The Republica Times &  Polarization, Propaganda & Video Game & Simulation & News Articles & Bad Actor, Good Actor & \cite{Rev2_2023_EvaluatingAsTools, Rev3_2024_LiteracyAtPlay} \\
        Post Facto &  (Mis)Information Literacy & Video Game & Choice-based Simulation, Trivia & News Articles & Good Actor & \cite{Rev2_2023_EvaluatingAsTools, Rev3_2024_LiteracyAtPlay} \\
        The Westport Independent & Propaganda, (Mis)Information Literacy & Video Game & Simulation & News Articles & Bad Actor, Good Actor & \cite{Rev3_2024_LiteracyAtPlay} \\
        NewsFeed Defenders & (Mis)Information Literacy & Video Game & Simulation & Social Media & Good Actor & \cite{Rev4_2023_PlayingFakeState, Rev2_2023_EvaluatingAsTools, Rev3_2024_LiteracyAtPlay}\\
        Fake-News-App:Vorsicht Giftstoffe im Handy! &  (Mis)Information Literacy & Video Game & Choice-based Simulation, Trivia & Multimodal (e.g., Video, Image, Audio), Social Media, News Articles & Good Actor & -- \\
        SWR FakeFinder Kids &  (Mis)Information Literacy, Trolls, Satire, Humor, Ads & Video Game & Trivia & Multimodal (e.g., Video, Image, Audio), Other, Social Media & Detached Evaluator & -- \\
        Ezra \& Alex &  (Mis)Information Literacy & Video Game & Point and Click & Multimodal (e.g., Video, Image, Audio) & Good Actor & --  \\ 
        SchlaWiener & (Mis)Information Literacy, Conspiracy Theories, Manipulation Strategies & Video Game & Decision Tree, Choice-based Simulation & Social Media, News Articles & Good Actor, Bad Actor & -- \\
        Escape Game & (Mis)Information Literacy & Video Game & Other, Trivia & Multimodal (e.g., Video, Image, Audio), News Articles & Good Actor & --\\
        Real, LOLZ, oops or fake & (Mis)Information Literacy & Video Game & Trivia & News Articles, Multimodal (e.g., Video, Image, Audio) & Detached Evaluator & \cite{Rev4_2023_PlayingFakeState} \\
        Headliner: Novinews & (Mis)Information Literacy, Propaganda & Video Game & Simulation & News Articles & Bad Actor, Good Actor & \cite{Rev2_2023_EvaluatingAsTools} \\
        Not for Broadcast &  Propaganda & Video Game & Simulation & News Articles & Good Actor, Bad Actor & -- \\
        Doubt is our Product &  (Mis)Information Literacy & Board Game & Board Game & Other & Good Actor, Bad Actor & -- \\
        Fake Hunter &  (Mis)Information Literacy& Video Game & Simulation & News Articles & Good Actor & -- \\
        Fight Fakes &  (Mis)Information Literacy & Classroom Game & Classroom & News Articles & Detached Evaluator & --\\
        Exit the Fake &  (Mis)Information Literacy & Classroom Game, Escape Room & Other & Multimodal (e.g., Video, Image, Audio), Social Media & Good Actor & -- \\
        Schnitzeljagd: Fake News &  (Mis)Information Literacy & Classroom Game, Video Game & Other, Trivia & News Articles, Social Media & Detached Evaluator & -- \\
        Fake-O-Mat & (Mis)Information Literacy, Disinformation, Polarization & Classroom Game, Tabletop Roleplay Game & RPG & News Articles & Good Actor & -- \\
        Choose your own fake news &  (Mis)Information Literacy, Trolls, Satire, Humor, Ads & Video Game & Decision Tree & Social Media, Other & Good Actor, Detached Evaluator & \cite{Rev4_2023_PlayingFakeState, Rev2_2023_EvaluatingAsTools} \\
        Spot The Troll &  Bots, Disinformation, (Mis)Information Literacy, Social Media & Video Game & Trivia & Social Media & Detached Evaluator & \cite{Rev2_2023_EvaluatingAsTools, Rev4_2023_PlayingFakeState}\\
        Influence Inc. &  (Mis)Information Literacy, Disinformation, Social Media& Video Game & Simulation & Social Media & Bad Actor & \cite{Rev2_2023_EvaluatingAsTools}\\
        No Place for the Dissident &  Polarization, Propaganda, Disinformation, Manipulation Strategies & Video Game & Simulation& Other & Bad Actor & \cite{Rev2_2023_EvaluatingAsTools} \\
        \bottomrule
    \end{longtable}
    }

\section{Search Queries for Academic Papers}

\begin{table*}[h]
\caption{Google Scholar Search Queries}
\label{tab:scholar-queries}
\small
\begin{tabular}{p{0.1\textwidth} p{0.89\textwidth}}
\toprule
\textbf{ID} & \textbf{Search Query String} \\
\midrule
\textbf{S1} & (misinformation OR "fake news" OR disinformation OR propaganda OR "information disorders") AND (education OR intervention OR pedagogy OR curriculum OR "media education" OR "public pedagogy") \\
\addlinespace
\textbf{S2} & ("media bias" OR framing OR "framing effects" OR "agenda-setting" OR priming OR "cognitive bias") AND (mitigation OR intervention OR "inoculation theory" OR "cognitive inoculation" OR resilience) \\
\addlinespace
\textbf{S3} & ("media literacy" OR "digital literacy" OR "news literacy" OR "information literacy" OR "social media literacy") AND (gamification OR "game-based learning" OR "serious games" OR "educational games" OR "interactive learning") \\
\addlinespace
\textbf{S4} & ("generative AI" OR GenAI OR "AI literacy" OR "AI-generated misinformation" OR deepfakes OR "synthetic media") AND (education OR "digital literacy" OR intervention OR "awareness raising") \\
\addlinespace
\textbf{S5} & ("misinformation video" OR "video literacy" OR "digital storytelling" OR "YouTube literacy" OR "TikTok literacy") AND (education OR intervention OR "edutainment" OR "instructional video" OR "media campaign") \\
\addlinespace
\textbf{S6} & ("critical thinking" OR "critical analysis" OR "civic online reasoning" OR OCR OR "online reasoning strategies") AND (education OR intervention OR "instructional design" OR learning) \\
\addlinespace
\textbf{S7} & (Misinformation OR ``Misinformation Video'' OR Media Bias OR Fake News OR ``Fake News Video'' OR ``Online Civic Reasoning'' OR ``News Media Literacy'') AND ( Intervention OR Mitigation OR Game OR Gamification OR Gamified) \\
\bottomrule
\end{tabular}
\end{table*}

\section{Additional Figures}
\begin{figure*}[h]
            \centering
            \includegraphics[width=0.7\linewidth]{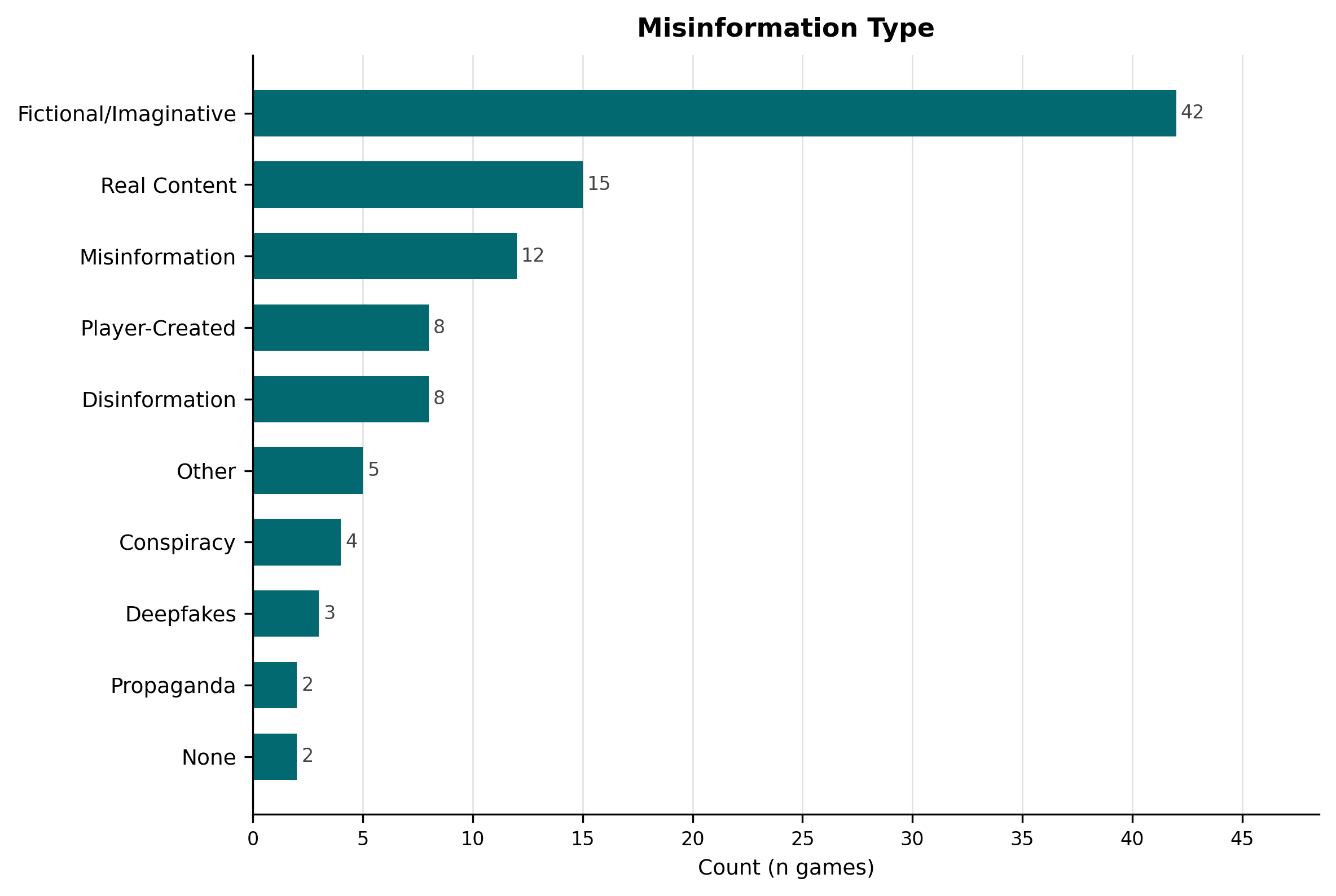}
            \caption{Occurrence of labels for misinformation type.}
            \Description{Horizontal bar chart showing how many games games carry each label (x-axis: count of games, y-axis: label). Fictional/Imaginative content has the highest count with 42,followed by real content (15), misinformation (12), player-created (8), disinformation (8), other (5), conspiracy (4), deepfakes (3), propaganda (2), and none (2). }
            \label{fig:misinfolabel}
        \end{figure*}

        \begin{figure*}[h]
            \centering
            \includegraphics[width=0.7\linewidth]{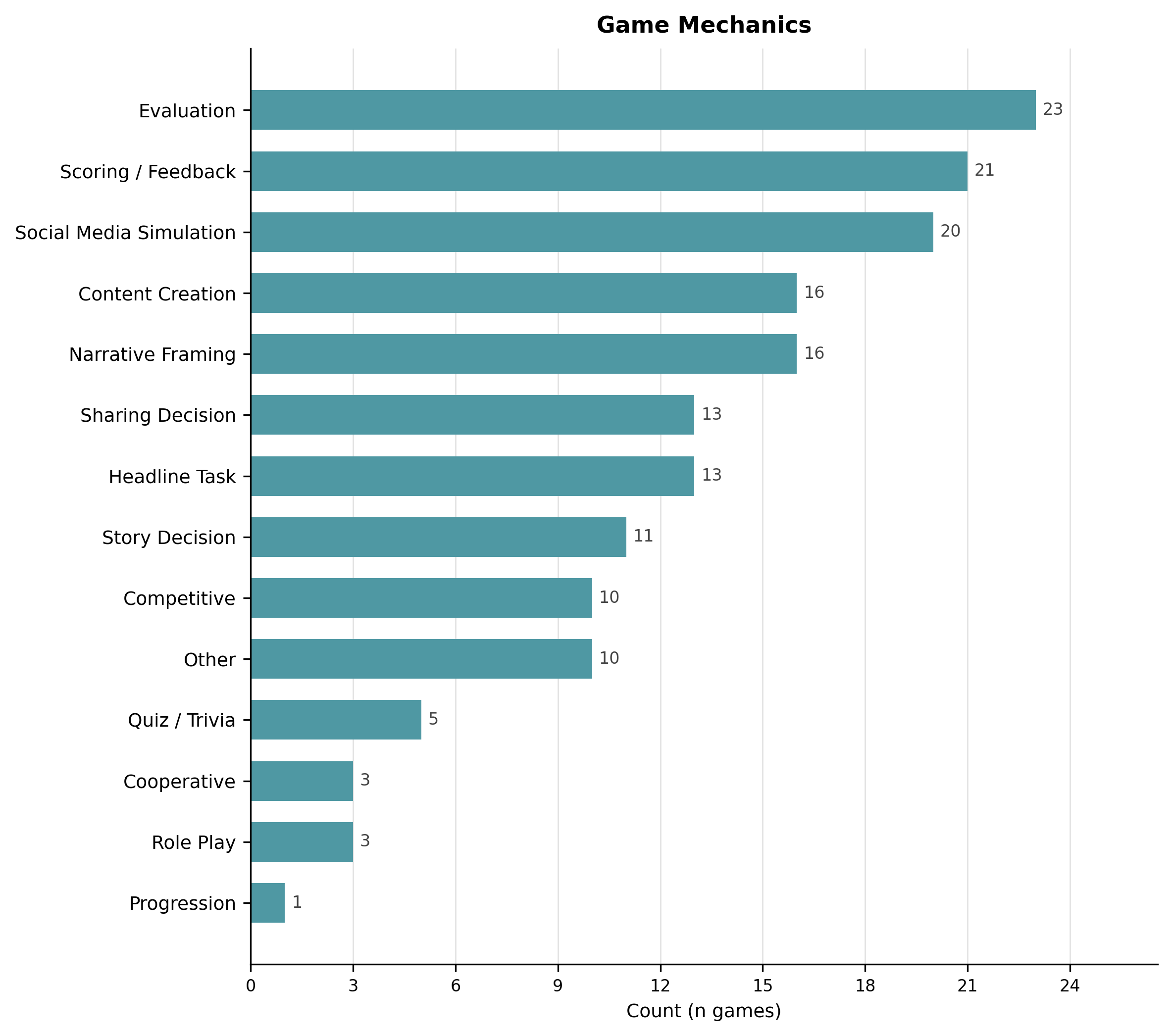}
            \caption{Occurrence of labels for game mechanics.}
            \Description{Horizontal bar chart showing how many games use each game mechanic. The most common are evaluation (23), scoring/feedback (21), and social-media simulation (20), followed by content creation (16), narrative framing (16), sharing decision (13), headline task (13), story decision (11), competitive (10), and other (10); quiz/trivia (5), cooperative (3), role-play (3), and progression (1) are least common.}
            \label{fig:mechanicslabel}
        \end{figure*}

        \begin{figure*}[h]
            \centering
            \includegraphics[width=0.98\linewidth]{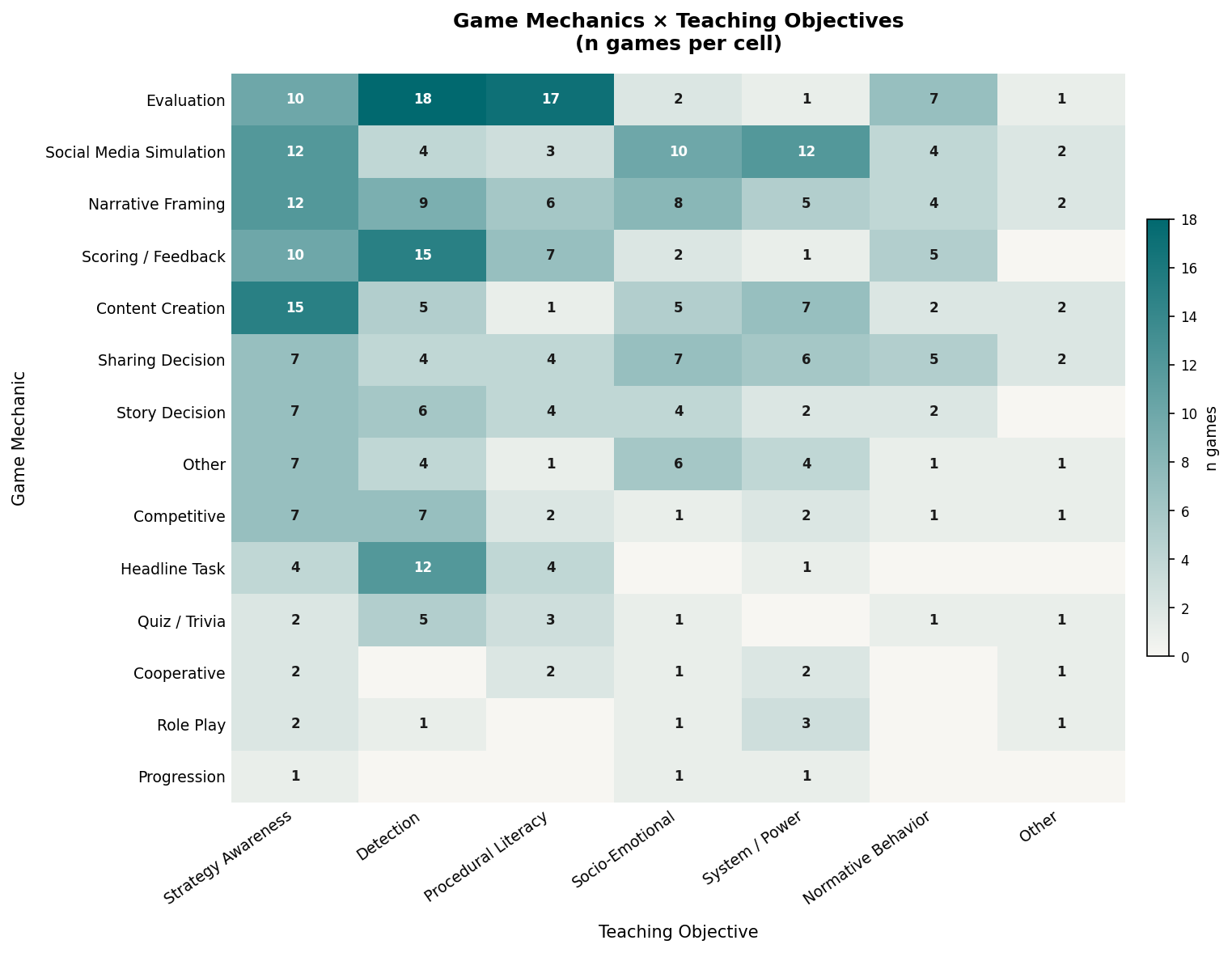}
            \caption{Co-occurrence of labels for teaching objective and game mechanic.}
            \Description{A heatmap titled "Game Mechanics × Teaching Objectives (n games per cell)" cross-tabulates 14 game mechanics (y-axis) against 7 teaching objectives (x-axis), with each cell showing the count of games that combine that mechanic with that objective. Cell color ranges from near-white (0 games) to dark teal (18 games). Teaching objectives (x-axis, left to right): Strategy Awareness, Detection, Procedural Literacy, Socio-Emotional, System/Power, Normative Behavior, Other.mGame mechanics (y-axis, top to bottom): Evaluation, Social Media Simulation, Narrative Framing, Scoring/Feedback, Content Creation, Sharing Decision, Story Decision, Other, Competitive, Headline Task, Quiz/Trivia, Cooperative, role-play, Progression. Dominant patterns: The three highest cells are Evaluation × Detection (18), Evaluation × Procedural Literacy (17), and Content Creation × Strategy Awareness (15), all appearing in deep teal. Scoring/Feedback × Detection (15), Headline Task × Detection (12), and Social Media Simulation × Socio-Emotional (10) and × System/Power (12) are also notably high. Detection and Strategy Awareness are the most densely covered teaching objectives overall, with high counts across many mechanics. Socio-Emotional and System/Power objectives are most strongly served by Social Media Simulation. Lower-frequency mechanics such as Cooperative, role-play, and Progression show sparse coverage across all objectives, with most cells at 0–2. The "Other" teaching objective column and the Normative Behavior column are sparsely populated across all mechanics, rarely exceeding 5.}
            \label{fig:heatmap}
        \end{figure*}
        
\begin{figure*}[h]
            \centering
            \includegraphics[width=0.8\linewidth]{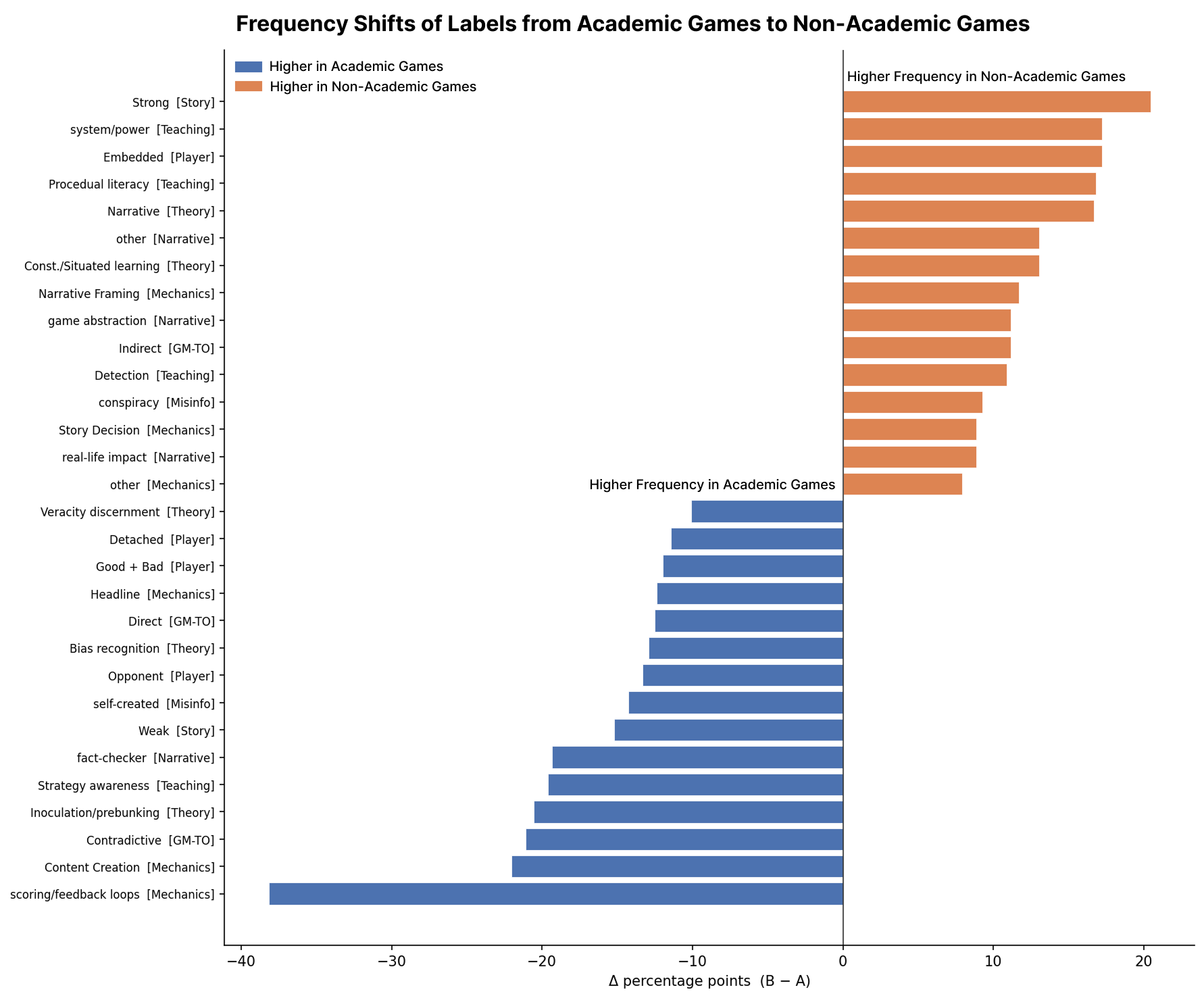}
            \caption{Differences between Academic and Non-Academic Games. Brackets behind the codes indicate the code category (Story Strength, Teaching Objective, Player Role, Teaching Theory, Game Mechanics, Narrative Setting, Type of Misinformation.). GM-TO stands for the relationship between teaching objective (TO) and the game mechanics (GM). The biggest difference are that non-academic games often have stronger stories while academic games rely on simple scoring and feedback mechanics, combined with misinformation creation for inoculation theory.}
            \Description{Diverging horizontal bar chart plotting the difference in percentage points (non-academic minus academic) for each coded label, colored by direction. Labels more frequent in non-academic games include strong story (the largest positive shift), system/power teaching, embedded player role, procedural-literacy and narrative theory, and situated/constructivist learning. Labels more frequent in academic games include scoring/feedback loops (the largest shift), content creation, contradiction, inoculation/prebunking, strategy awareness, veracity discernment, and detached or opponent player roles. Bracketed tags mark each label's category (Story, Teaching, Player, Theory, Mechanics, Narrative, GM-TO).}
            \label{fig:academicvsnonacademic}
        \end{figure*}

\clearpage
\onecolumn
\hypertarget{annotation}{}
\pagestyle{empty}
\lstset{
  basicstyle=\footnotesize\ttfamily,
  breaklines=true,
  breakatwhitespace=false,
  columns=flexible,
  numbers=none
}

\definecolor{Primary}{RGB}{59, 130, 246}    
\definecolor{PrimaryDark}{RGB}{30, 64, 175} 
\definecolor{LightBg}{RGB}{239, 246, 255}   
\definecolor{TextDark}{RGB}{31, 41, 55}     
\definecolor{TextMuted}{RGB}{107, 114, 128} 

\begin{tikzpicture}[remember picture, overlay]
  \fill[Primary] ([xshift=0cm,yshift=0cm]current page.north west) rectangle ([xshift=\paperwidth,yshift=-0.4cm]current page.north west);
\end{tikzpicture}

\vspace{0.8cm}
\begin{center}
  {\fontsize{22}{26}\selectfont\sffamily\bfseries \textcolor{PrimaryDark}{CiteAssist}}\\[0.2em]
  {\Large\sffamily\scshape \textcolor{TextMuted}{Citation Sheet}}\\[0.8em]
  {\small\sffamily Generated with \href{https://citeassist.uni-goettingen.de/}{\textcolor{Primary}{\texttt{citeassist.uni-goettingen.de}}}
  \CiteAssistCite{}
  }\end{center}

\begin{center}
\vspace{1em}
\begin{tikzpicture}
\draw[Primary, line width=0.6pt] (0,0) -- (\textwidth,0);
\end{tikzpicture}
\vspace{1.2em}
\end{center}

\begin{tcolorbox}[enhanced,
                 frame hidden,
                 boxrule=0pt,
                 borderline west={2pt}{0pt}{Primary},
                 colback=LightBg,
                 sharp corners,
                 breakable,
                 fonttitle=\sffamily\bfseries\large,
                 coltitle=Primary,
                 title=BibTeX Entry,
                 attach title to upper={\vspace{0.2em}\par},
                 left=12pt]
\lstset{
    inputencoding = utf8,  
    extendedchars = true,  
    literate      =        
      {á}{{\'a}}1  {é}{{\'e}}1  {í}{{\'i}}1 {ó}{{\'o}}1  {ú}{{\'u}}1
      {Á}{{\'A}}1  {É}{{\'E}}1  {Í}{{\'I}}1 {Ó}{{\'O}}1  {Ú}{{\'U}}1
      {à}{{\`a}}1  {è}{{\`e}}1  {ì}{{\`i}}1 {ò}{{\`o}}1  {ù}{{\`u}}1
      {À}{{\`A}}1  {È}{{\`E}}1  {Ì}{{\`I}}1 {Ò}{{\`O}}1  {Ù}{{\`U}}1
      {ä}{{\"a}}1  {ë}{{\"e}}1  {ï}{{\"i}}1 {ö}{{\"o}}1  {ü}{{\"u}}1
      {Ä}{{\"A}}1  {Ë}{{\"E}}1  {Ï}{{\"I}}1 {Ö}{{\"O}}1  {Ü}{{\"U}}1
      {â}{{\^a}}1  {ê}{{\^e}}1  {î}{{\^i}}1 {ô}{{\^o}}1  {û}{{\^u}}1
      {Â}{{\^A}}1  {Ê}{{\^E}}1  {Î}{{\^I}}1 {Ô}{{\^O}}1  {Û}{{\^U}}1
      {œ}{{\oe}}1  {Œ}{{\OE}}1  {æ}{{\ae}}1 {Æ}{{\AE}}1  {ß}{{\ss}}1
      {ẞ}{{\SS}}1  {ç}{{\c{c}}}1 {Ç}{{\c{C}}}1 {ø}{{\o}}1  {Ø}{{\O}}1
      {å}{{\aa}}1  {Å}{{\AA}}1  {ã}{{\~a}}1  {õ}{{\~o}}1 {Ã}{{\~A}}1
      {Õ}{{\~O}}1  {ñ}{{\~n}}1  {Ñ}{{\~N}}1  {¿}{{?\`}}1  {¡}{{!\`}}1
      {„}{\quotedblbase}1 {“}{\textquotedblleft}1 {–}{$-$}1
      {°}{{\textdegree}}1 {º}{{\textordmasculine}}1 {ª}{{\textordfeminine}}1
      {£}{{\pounds}}1  {©}{{\copyright}}1  {®}{{\textregistered}}1
      {«}{{\guillemotleft}}1  {»}{{\guillemotright}}1  {Ð}{{\DH}}1  {ð}{{\dh}}1
      {Ý}{{\'Y}}1    {ý}{{\'y}}1    {Þ}{{\TH}}1    {þ}{{\th}}1    {Ă}{{\u{A}}}1
      {ă}{{\u{a}}}1  {Ą}{{\k{A}}}1  {ą}{{\k{a}}}1  {Ć}{{\'C}}1    {ć}{{\'c}}1
      {Č}{{\v{C}}}1  {č}{{\v{c}}}1  {Ď}{{\v{D}}}1  {ď}{{\v{d}}}1  {Đ}{{\DJ}}1
      {đ}{{\dj}}1    {Ė}{{\.{E}}}1  {ė}{{\.{e}}}1  {Ę}{{\k{E}}}1  {ę}{{\k{e}}}1
      {Ě}{{\v{E}}}1  {ě}{{\v{e}}}1  {Ğ}{{\u{G}}}1  {ğ}{{\u{g}}}1  {Ĩ}{{\~I}}1
      {ĩ}{{\~\i}}1   {Į}{{\k{I}}}1  {į}{{\k{i}}}1  {İ}{{\.{I}}}1  {ı}{{\i}}1
      {Ĺ}{{\'L}}1    {ĺ}{{\'l}}1    {Ľ}{{\v{L}}}1  {ľ}{{\v{l}}}1  {Ł}{{\L{}}}1
      {ł}{{\l{}}}1   {Ń}{{\'N}}1    {ń}{{\'n}}1    {Ň}{{\v{N}}}1  {ň}{{\v{n}}}1
      {Ő}{{\H{O}}}1  {ő}{{\H{o}}}1  {Ŕ}{{\'{R}}}1  {ŕ}{{\'{r}}}1  {Ř}{{\v{R}}}1
      {ř}{{\v{r}}}1  {Ś}{{\'S}}1    {ś}{{\'s}}1    {Ş}{{\c{S}}}1  {ş}{{\c{s}}}1
      {Š}{{\v{S}}}1  {š}{{\v{s}}}1  {Ť}{{\v{T}}}1  {ť}{{\v{t}}}1  {Ũ}{{\~U}}1
      {ũ}{{\~u}}1    {Ū}{{\={U}}}1  {ū}{{\={u}}}1  {Ů}{{\r{U}}}1  {ů}{{\r{u}}}1
      {Ű}{{\H{U}}}1  {ű}{{\H{u}}}1  {Ų}{{\k{U}}}1  {ų}{{\k{u}}}1  {Ź}{{\'Z}}1
      {ź}{{\'z}}1    {Ż}{{\.Z}}1    {ż}{{\.z}}1    {Ž}{{\v{Z}}}1  {ž}{{\v{z}}}1
  }
\begin{lstlisting}
@article{abedLandscapeMisinformationGames2026,
  author={Abed, Omed and Hinterreiter, Smi and Guo, Sijia and Echizen, Isao and Spinde, Timo and Grosse-Kampmann, Matteo},
  chapter={Article GAMES068 [Conditionally Accepted]},
  copyright={CC-By Attribution 4.0 International},
  journal={Proc. ACM Hum.-Comput. Interact.},
  month={nov},
  number={7},
  pages={38},
  publisher={ACM},
  shorttitle={The {Landscape} of {Misinformation} {Literacy} {Games}},
  title={The {Landscape} of {Misinformation} {Literacy} {Games}: {A} {Systematic} {Mapping} {Review} of {Game} {Designs} and {Evaluation} {Practices}},
  volume={10},
  year={2026},
  note = {Preprint of Conditionally Accepted Manuscript.},
  homepage = {https://media-bias-group.github.io/Project-Page-Misinformation-Games-Review/}
}
\end{lstlisting}
\end{tcolorbox}

\vfill
\begin{tikzpicture}
\draw[Primary!40, line width=0.4pt] (0,0) -- (\textwidth,0);
\end{tikzpicture}
\begin{center}
\small\sffamily\textcolor{TextMuted}{Generated \today}
\end{center}

\end{document}